\documentclass[english,15pt]{article}
\usepackage[T1]{fontenc}
\usepackage[latin1]{inputenc}
\usepackage{geometry}
\usepackage{float}
\usepackage{mathrsfs}
\usepackage{amsmath}
\usepackage{amssymb}
\usepackage{comment}

\usepackage{graphicx}
\usepackage[pagebackref=true]{hyperref}
\hypersetup{
    colorlinks=true,
    linkcolor=blue,
    filecolor=magenta,      
    urlcolor=cyan,
    citecolor = blue,
}

\usepackage{todonotes}

\makeatletter

\floatstyle{ruled}
\usepackage{algorithm}
 \usepackage{algorithmicx}

\usepackage{amsthm}

\usepackage{mathrsfs}

\usepackage{amsfonts}

\usepackage{epsfig}

\usepackage{bm}

\usepackage{mathrsfs}

\usepackage{enumerate}

\@ifundefined{definecolor}{\@ifundefined{definecolor}
 {\@ifundefined{definecolor}
 {\usepackage{color}}{}
}{}
}{}

\usepackage{subfig}\usepackage[all]{xy}

\newtheorem{theorem}{Theorem}[section]
\newtheorem{lem}{Lemma}[section]
\newtheorem{rem}{Remark}[section]
\newtheorem{prop}{Proposition}[section]

\newcounter{hypA}

\newcounter{hypB}

\newcounter{hypD}

\usepackage{babel}\date{}

\newcommand{\Pa}{ {\cal P }}

\numberwithin{equation}{section}

\makeatother

\begin{document}

\begin{center}

{\Large \textbf{Unbiased Estimation of the Vanilla and  Deterministic \\ Ensemble  Kalman--Bucy Filters}}

\vspace{0.5cm}

MIGUEL \'{A}LVAREZ, NEIL K. CHADA \& AJAY JASRA

{\footnotesize Applied Mathematics and Computational Science Program \\ Computer, Electrical and Mathematical Sciences and Engineering Division, \\ King Abdullah University of Science and Technology, Thuwal, 23955, KSA.} \\
{\footnotesize E-Mail:\,} \texttt{\emph{\footnotesize miguelangel.alvarezballesteros@kaust.edu.sa, neil.chada@kaust.edu.sa, \\ ajay.jasra@kaust.edu.sa}} \\
\begin{abstract}
In this article we consider the development of an unbiased estimator for the ensemble Kalman--Bucy filter (EnKBF). The EnKBF is a continuous-time
filtering methodology which can be viewed as a continuous-time analogue of the famous discrete-time ensemble Kalman filter. Our unbiased estimators will be motivated from
recent work [Rhee \& Glynn 2010, \cite{RG15}] which introduces randomization as a means to produce unbiased and finite variance estimators. The randomization enters through both the level of discretization, and through the number of samples at each level. Our estimator will be specific to linear and
Gaussian settings, where we know that the EnKBF is consistent, in the particle limit $N \rightarrow \infty$, with the KBF. We highlight this for two particular
variants of the EnKBF, i.e. the deterministic and vanilla variants, and demonstrate this on a linear Ornstein--Uhlenbeck process. We compare this with the EnKBF 
and the multilevel (MLEnKBF), for experiments with varying dimension size. We also provide a proof of the multilevel deterministic EnKBF, which provides a guideline for some of the unbiased methods.
\end{abstract}

\end{center}

\section{Introduction}
Let $(\Omega,\mathcal{F},\mathbb{P})$ denote a probability space, such that $\mathbb{P}$ is a probability measure defined on the space  $(\Omega,\mathcal{F})$.
For this article we consider two stochastic processes $\{Y_t\}_{t \geq 0}$ and $\{X_t\}_{t \geq 0}$ defined on $\mathbb{P}$, which are specified through the following
stochastic differential equations.
\begin{align}
\label{eq:dat}
dY_t &=  h(X_t) dt +  dV_t,   \\
\label{eq:sig}
dX_t &=   f(X_t) dt +  \sigma(X_t) dW_t, 
\end{align}
 with $h:\mathbb{R}^{d_x} \rightarrow \mathbb{R}^{d_y}$ and $f:\mathbb{R}^{d_x} \rightarrow \mathbb{R}^{d_x}$ denoting potentially nonlinear functions, $\sigma: \mathbb{R}^{d_x} \rightarrow  \mathbb{R}^{d_x \times d_x}$
 denoting the diffusion coefficient and $\{V_t\}_{t \geq 0},\{W_t\}_{t \geq 0}$ are independent $d_y-$ and $d_x-$dimensional Brownian motions. Our motivation from this setup is in the filtering
 problem \cite{BC09,CR11,AJ70}, which is concerned with the recursive estimation of the hidden signal process $\{X_t\}_{t \geq 0}$ through the use of noisy observed process $\{Y_t\}_{t \geq 0}$. The 
 filtering problem is of particular interest, which is demonstrated through its application in numerous fields. Most notably they include, but not limited to, atmospheric sciences, numerical
 weather prediction, mathematical finance and more recently machine learning \cite{DP10,SAA17, MW06, ORL08}. 
 Mathematically the aim of the filtering problem is to compute   conditional expectation $\mathbb{E}[\varphi(X_t)|\mathscr{F}_t]$, 
 where $\varphi:\mathbb{R}^{d_x} \rightarrow \mathbb{R}$  is an appropriately integrable function and $\{\mathscr{F}_t\}_{t \geq 0}$ is the filtration generated by the observed process \eqref{eq:dat}.
 To solve the filtering problem analytically it is often challenging and intractable, therefore instead one must compute approximate solutions, which are done through filtering methods. Commonly
 filtering methods can be split into two categories, the first being particle filters, or sequential Monte Carlo \cite{CP20,PD04}, which use resampling strategies and importance sampling. The later which will be the focus of this work are Kalman-based methods.
 
 Kalman filtering dates back to the original work of Rudolph Kalman, who proposed a way to perform filtering in a linear and Gaussian setting. The proposed methodology was the Kalman filter (KF) which updates a state assumed to be Gaussian, based on updating the mean and covariance. The filter is significant as it is 
 the optimal filter in the linear and Gaussian setting, i.e. it minimizes the mean square error of the approximation to exact solution. Since its formulation it has seen significant gains in new methodologies and applications, in particular the ensemble Kalman filter (EnKF) \cite{GE09,GE94}. The EnKF is a Monte Carlo version of the KF which updates an ensemble of particles based on the sample mean and covariance. As a result of this modification, the EnKF is less computationally costly than the KF, as one does not require the update of the full covariance matrix at every iteration. Another direction has been the understanding of these methods in the continuous-time setting. For the KF its continuous analogue is the Kalman--Bucy filter (KBF) \cite{bucy,AJ70}, and similarly for the EnKF it is the ensemble Kalman--Bucy filter (EnKBF). Both continuous-time algorithms are challenging to work with, both to derive and analysis and practically. However there has been a growing interest in the EnKBF for both theory and applications. In particular the work of Del Moral and coauthors have aimed to analyze various theoretical properties \cite{BD17,DT18,DKT17}, while other directions have been on applications to non-linear filtering and uncertainty quantification \cite{CJY20,CDJ21,RCJ21,WRS18,WT19}.
 
With filtering methodologies, for both discrete and continuous-time problems, one must require a time-discretization in order to solve the problem. Such a discretization, as in the context of our work, induces a discretization bias. Therefore it is of interest to introduce debiased or unbiased estimators of the
conditional expectation $\mathbb{E}[\varphi(X_t)|\mathscr{F}_t]$. A recent and popular methodology for unbiased estimation is that of Rhee and Glynn \cite{GR14,RG15}. The underlying idea behind their methodology is that to attain unbiased estimation they  randomize over the time-discretizations.
 This involves placing a probability distribution over the level of time-discretization, and then to sample from that particular level.  For filtering this has been
 considered in the contexts of different works related to partially observed diffusions, discretized models and gradient estimates \cite{CFJ21,HHJ21,HJL21,JLY20,MV18}. However as of yet these techniques have not been applied to Kalman-based filtering,
  which thus acts as the motivation for this article.
 
 In particular we aim to developed an unbiased estimator for the EnKBF, based on the works of \cite{GR14,RG15}. This will be done for two of the unbiased 
 estimators introduced in \cite{RG15} which are the single-term estimator and the coupled sum estimator. In order to assess the performance of the 
 unbiased estimator, we compare this with the multilevel EnKBF (MLEnKBF) \cite{CJY20,RCJ21}. The multilevel approach \cite{MBG08,MBG15} is a variance reduction
 approach for Monte Carlo which has seen particular interest in the filtering community \cite{HLT16,JKL17}. By applying MLMC one can reduce the error-to-cost rates of the mean
 square error. We will use these same plots in order to compare both. This will be tested on a linear Ornstein--Uhlenbeck process, where we compare our unbiased estimators of the EnKBF to two particular variants of the MLEnKBF. \textcolor{black}{To complement this work, we also derive MLMC rates, for the deterministic EnKBF, which is an extension of the work of \cite{CJY20}}, whose results were specific to the vanilla version.  We will conduct this on two different dimensional examples, i.e. $d_x=d_y=2$ and 
 $d_x=d_y=500$. From our results we will demonstrate in order to attain an MSE of order $\mathcal{O}(\epsilon^2)$ we require a computational cost 
 of  $\mathcal{O}(\epsilon^{-2}\log(\epsilon)^{2+\beta})$, for $\beta>0$ compared to $\mathcal{O}(\epsilon^{-2}\log(\epsilon)^2)$ of the MLEnKBF.  
 We emphasis with this work our focus is not on proving our estimator is unbiased, with finite variance, as it beyond the scope of this paper, which we leave for future work.
 
 The outline of this paper is as follows: in  \autoref{sec:model} we begin with a review on the KBF and EnKBF. We then discuss the MLMC and its application to the EnKBF, resulting in the MLEnKBF.  \autoref{sec:unb} will introduce the unbiased estimator of the EnKBF. Our numerical experiments are conducted in  \autoref{sec:num} comparing our two unbiased estimators to that of the MLEnKBFs. We conclude our work with final remarks, and future directions of work in \autoref{sec:conc}.
 The proof of our main result will be provided in the Appendix.

\section{Model and Background}
\label{sec:model}

In this section we review the key concepts behind stochastic filtering, while introducing the necessary material on the Kalman--Bucy filters and various ensemble Kalman--Bucy filters (EnKBF).
We will also provide an overview on the underlying idea of multilevel Monte Carlo, before discussing and introducing the multilevel EnKBF. Finally we state our main result 
from the paper, related to the complexity analysis of the deterministic multilevel EnKBF.

\subsection{Kalman--Bucy Filters}

Our motivation from this work is related to the linear filtering problem, 
\begin{align}
\label{eq:data}
dY_t & =  CX_t dt + R_2^{1/2} dV_t, \\
\label{eq:signal}
dX_t & =  A X_t dt + R_1^{1/2} dW_t,
\end{align}
where $(Y_t,X_t)\in\mathbb{R}^{d_y}\times\mathbb{R}^{d_x}$, $(V_t,W_t)$ is a $(d_y+d_x)-$dimensional standard Brownian motion,
$A$ is a square $d_x\times d_x$ matrix, $C$ is a $d_y\times d_x$ matrix, $Y_0=0$, $X_0\sim\mathcal{N}_{d_x}(\mathcal{M}_0,\mathcal{P}_0)$
($d_x-$dimensional Gaussian distribution, mean $\mathcal{M}_0$, covariance matrix $\mathcal{P}_0$)
and $R_1^{1/2},R_2^{1/2}$ are square (of the appropriate dimension), symmetric and invertible matrices. It is well-known that, letting
$\{\mathscr{F}_t\}_{t\geq 0}$ be the filtration generated by the observations, the conditional probability of $X_t$ given $\mathscr{F}_t$ is a
Gaussian distribution with mean and covariance matrix 
$$
\mathcal{M}_t := \mathbb{E}[X_t|\mathscr{F}_t], \quad \mathcal{P}_t := \mathbb{E}\Big([X_t - \mathbb{E}(X_t|\mathscr{F}_t)][X_t - \mathbb{E}(X_t|\mathscr{F}_t)]^{\top}\Big),
$$
given by the Kalman--Bucy and Ricatti equations \cite{DT18}
\begin{align}
\label{eq:kbf}
d\mathcal{M}_t &= A \mathcal{M}_t dt + \mathcal{P}_tC^{\top}R^{-1}_2\Big(dY_t - C\hat{X_t}dt\Big), \\
\label{eq:ricc}
\partial_t\mathcal{P}_t &= \textrm{Ricc}(\mathcal{P}_t),
\end{align}
where the Riccati drift is defined as
$$
\textrm{Ricc}(Q) = AQ + QA^{\top}-QSQ + R, \quad \textrm{with} \ R =R_1 \quad \textrm{and} \ S:=C^{\top}R^{-1}_2C.
$$
 A derivation of \eqref{eq:kbf} - \eqref{eq:ricc} can be found in \cite{AJ70}.

The Kalman--Bucy diffusion is a conditional McKean-Vlasov diffusion process (e.g.~\cite{PD04}), where for this article we work with 
two different processes of the form
\begin{align}
\label{eq:non-lin}
d\overline{X}_t &= A\overline{X}_t dt + R_1^{1/2}d\overline{W}_t + \mathcal{P}_{\eta_t}C^{\top}R_2^{-1}\Big(dY_t -
\Big[C\overline{X}_tdt+R_2^{1/2}d\overline{V}_t\Big]\Big), \\
d\overline{X}_t & = A\overline{X}_t~dt+R^{1/2}_{1}~d\overline{W}_t+\Pa_{\eta_t}C^{\top}R^{-1}_{2}~\left[dY_t-\left(\frac{1}{2}C\left[\overline{X}_t+\eta_t(e)\right]dt\right)\right],
\label{eq:determ_kbf}
\end{align}
where $(\overline{V}_t,\overline{W}_t,\overline{X}_0)$ are independent copies of $(V_t,W_t,X_0)$ and covariance
$$
\mathcal{P}_{\eta_t} = \eta_t\Big([e-\eta_t(e)][e-\eta_t(e)]^{\top}\Big), \quad \eta_t:= \mathrm{Law}(\overline{X}_t|\mathscr{F}_t),
$$
such that $\eta_t$ is the conditional law of $\overline{X}_t$ given $\mathscr{F}_t$ and $e(x)=x$.  We will explain the difference of each diffusion process \eqref{eq:non-lin} - \eqref{eq:determ_kbf} in succeeding subsections.
It is important to note that the nonlinearity  in \eqref{eq:non-lin}-\eqref{eq:determ_kbf} does not depend on the distribution of the state $\mathrm{Law}(\overline{X}_t)$ but on the conditional distribution $\eta_t$, and $\mathcal{P}_{\eta_t}$ alone does not depend on $\mathscr{F}_t$.
It is known that the conditional  expectations of the random states $\overline{X}_t$ and their conditional covariance matrices $\mathcal{P}_{\eta_t}$, w.r.t. $\mathscr{F}_t$, satisfy the Kalman--Bucy and the Riccati equations. In addition, for any $t\in\mathbb{R}^+$
$$
\eta_t:= \mathrm{Law}(\overline{X}_t|\mathscr{F}_t) = \mathrm{Law}({X}_t|\mathscr{F}_t).
$$
As a result, an alternative to recursively computing \eqref{eq:kbf} - \eqref{eq:ricc}, is to generate $N$ i.i.d.~samples from 
the Kalman--Bucy diffusions and apply a Monte Carlo approximation.

\subsection{Ensemble Kalman--Bucy Filter}

Exact simulation from \eqref{eq:non-lin} is typically not possible, as one cannot compute $\mathcal{P}_{\eta_t}$ exactly. 
The ensemble Kalman--Bucy filter (EnKBF) can be used to deal with this issue. The EnKBF coincides with the mean-field particle interpretation of the diffusion \eqref{eq:non-lin}. The EnKBF is an $N-$particle system that is simulated as follows, for the $i^{th}-$particle, $i\in\{1,\dots,N\}$: 
\begin{align}
\label{eq:enkbf}
d\xi_t^i &= A\xi_t^i dt + R_1^{1/2} d\overline{W}_t^i + {P}_t^NC^{\top}R_2^{-1}\Big(dY_t -
\Big[C\xi_t^idt+R_2^{1/2}d\overline{V}_t^i\Big]\Big), \\
d\xi_t^i & = A~\xi_t^i~dt~+~R^{1/2}_{1}~d\overline{W}^i_t+P^N_tC^{\top}R^{-1}_{2}~\left[dY_t-\left(\frac{1}{2}C\left[\xi_t^i+ m_t^N \right]dt\right)\right],
\label{eq:denkbf}
\end{align}
such that
\begin{align*}
{P}_t^N & =  \frac{1}{N -1}\sum_{i=1}^N (\xi_t^i-m^N_t)(\xi_t^i-m^N_t)^{\top},\\
m^N_t & =  \frac{1}{N }\sum_{i=1}^N \xi_t^i,
\end{align*}
where $\xi_0^i\stackrel{\textrm{i.i.d.}}{\sim}\mathcal{N}_{d_x}(\mathcal{M}_0,\mathcal{P}_0)$. It is remarked that when $C=0$, \eqref{eq:enkbf} reduces to $N$ independent copies of an Ornstein--Uhlenbeck process. 

Above we have two particular variants of the EnKBF, which we will use throughout the article and test numerically. The first variant \eqref{eq:enkbf} is commonly referred
to as the vanilla EnKBF (VEnKBF) which is the most widely used variant in practice, and also regarding theory derived. What makes it unique is that it contains
perturbed observations. This is different to the other variant \eqref{eq:denkbf} which we refer to as the deterministic EnKBF (DEnKBF). This is because there are no perturbed
observations, therefore the only difference arises in the innovation function, i.e. the discrepancy between the data and solution evaluated in the operator $A \in \mathbb{R}^{d_x \times d_x}$. Related to this, there
exists another popular variant known as the deterministic transport EnKBF (DTEnKBF). However we omit such introducing it, as evidentially from \cite{RCJ21}, the numerics suggest
the behavior of the DTEnKBF is different.

In practice, one will not have access to an entire trajectory of observations. Thus numerically, one often works with a time discretization, such as the Euler method.
Let $\Delta_l=2^{-l}$ then we will generate the system for $(i,k)\in\{1,\dots,N\}\times\mathbb{N}_0=\mathbb{N}\cup\{0\}$ as
\begin{align}
\label{eq:enkf_ps}
\xi_{(k+1)\Delta_l}^i &= \xi_{k\Delta_l}^i + A\xi_{k\Delta_l}^i\Delta_l + R_1^{1/2} [\overline{W}_{(k+1)\Delta_l}^i-\overline{W}_{k\Delta_l}^i] \\
&+ P_{k\Delta_l}^NC^{\top}R_2^{-1}\Big([Y_{(k+1)\Delta_l}-Y_{k\Delta_l}]-
\Big[C\xi_{k\Delta_l}^i\Delta_l+R_2^{1/2}[\overline{V}_{(k+1)\Delta_l}^i-\overline{V}_{k\Delta_l}^i]\Big]\Big), \nonumber \\ \nonumber \\
    \label{eq:denkbf}
 \xi_{(k+1)\Delta_l}^i & =  \xi_{k\Delta_l}^i + A  \xi_{k\Delta_l}^i  \Delta_l + R_1^{1/2} \big\{\overline{W}_{(k+1)\Delta_l}^i-\overline{W}_{k\Delta_l}^i \big\} + \\
  &P^N_{k\Delta_l} C^{\top} R_2^{-1} \left( \big[Y_{(k+1)\Delta_l} - Y_{k\Delta_l} \big] - C \left[\dfrac{ \xi_{k\Delta_l}^i + m^N_{k\Delta_l}}{2} \right]\Delta_l \right), \nonumber
\end{align}
such that
\begin{align*}
P_{k\Delta_l}^N & =  \frac{1}{N -1}\sum_{i=1}^N (\xi_{k\Delta_l}^i-m_{k\Delta_l}^N)(\xi_{k\Delta_l}^i-m_{k\Delta_l}^N)^{\top}, \\
m_{k\Delta_l}^N & =  \frac{1}{N }\sum_{i=1}^N \xi_{k\Delta_l}^i,
\end{align*}
and $\xi_0^i\stackrel{\textrm{i.i.d.}}{\sim}\mathcal{N}_{d_x}(\mathcal{M}_0,\mathcal{P}_0)$. For $l\in\mathbb{N}_0$ given. Denote by $\eta_t^{N,l}$ as the $N-$empirical
probability measure of the particles $(\xi_t^1,\dots,\xi_t^N)$, where $t\in\{0,\Delta_l,2\Delta_l,\dots\}$. For $\varphi:\mathbb{R}^{d_x}\rightarrow\mathbb{R}^{d_x}$
we will use the notation 
\begin{equation}
\label{eq:eta}
\eta_t^{N,l}(\varphi):=\frac{1}{N}\sum_{i=1}^N\varphi(\xi_t^{i}).
\end{equation}

The EnKBF has seen recent progression in research, both methodologically and theoretically. For the former it has seen connections to particle filters,
which have demonstrated overcoming the curse of dimensionality known as the feedback particle filter \cite{SKP19}. Furthermore localization techniques have been
applied for nonlinear problems \cite{WRS18,WT19}. For the later, there has been significant gains in deriving analysis, such as continuous-time limits, stability bounds and
propagation of chaos results \cite{BD17,DT18,DKT17}. For a consise review on topics related to the EnKBF, we refer the reader to the work of Bishop et al. \cite{BD20}.

\subsection{Multilevel Monte Carlo}

To help facilitate our main contribution, we seek to compare our methodology of the unbiased EnKBF to its multilevel counterpart.
Before discussing the MLEnKBF, we begin with a short and concise introduction into multilevel Monte Carlo (MLMC). The idea of MLMC
was proposed and popularized by Giles and Hendrich \cite{MBG08,MBG15}, with initial applications to diffusion processes arising in mathematical finance,
which has since seen applications to filtering, uncertainty quantification and others.

Let $\pi$ be a probability on a measurable space $(\mathsf{X},\mathscr{X})$ and for $\pi-$integrable $\varphi:\mathsf{X}\rightarrow\mathbb{R}$
consider the problem of estimating $\pi(\varphi)=\mathbb{E}_{\pi}[\varphi(X)]$. We assume that we only have access to a sequence of approximations of $\pi$,  $\{\pi_l\}_{l\in\mathbb{N}_0}$, also
each defined on $(\mathsf{X},\mathscr{X})$ and we are now interested in estimating $\pi_l(\varphi)$, such that 
$\lim_{l\rightarrow\infty}|[\pi_l-\pi](\varphi)|=0$. 
Note that it is assumed that (for instance) the cost of simulation from $\pi_l$ increases with $l$, but, the approximation error between $\pi$ and $\pi_l$ is also falling as $l$ grows.
Then one can consider the telescoping sum
$$
\pi_L(\varphi) = \pi_0(\varphi) + \sum^L_{l=1}[\pi_l-\pi_{l-1}](\varphi).
$$
Then the idea of the algorithm is as follows:
\smallskip
\begin{enumerate}
\item{Approximate $\pi_0(\varphi)$ by using i.i.d.~sampling from $\pi_0$.}
\item{Independently for each $l\in\{1,\dots,L\}$ and the sampling in 1.~approximate $[\pi_l-\pi_{l-1}](\varphi)$ by i.i.d.~sampling from a coupling of $(\pi_l,\pi_{l-1})$.}
\end{enumerate}
The key point is to construct a coupling in 2.~so that the mean square error (MSE) can be reduced relative to i.i.d.~sampling from $\pi_L$. This latter construction often
relies on the specific properties of $(\pi_l,\pi_{l-1})$. 

Denoting $N_0\in\mathbb{N}$ i.i.d.~samples from $\pi_0$ as $(X^{1,0},\dots,X^{N_0,0})$ and for $l\in\{1,\dots,L\}$, $N_l\in\mathbb{N}$ samples
from a coupling of $(\pi_l,\pi_{l-1})$ as $((X^{1,l},\tilde{X}^{1,l-1}),\dots,(X^{N_l,l},\tilde{X}^{N_l,l-1}))$, one has the MLMC approximation of $\mathbb{E}_{\pi_L}[\varphi(X)]$
$$
\pi_L^{ML}(\varphi) := \frac{1}{N_0}\sum_{i=1}^{N_0}\varphi(X^{i,0}) + \sum_{l=1}^L \frac{1}{N_l}\sum_{i=1}^{N_l}\{\varphi(X^{i,l})-\varphi(\tilde{X}^{i,l-1})\}.
$$
The MSE is then
$$
\mathbb{E}[(\pi_L^{ML}(\varphi)-\pi(\varphi))^2] = \mathbb{V}\textrm{ar}[\pi_L^{ML}(\varphi)] + 
[\pi_L-\pi](\varphi)^2,
$$
where $\mathbb{V}\textrm{ar}[\cdot]$ denotes the variance (which we assume exists). One has
$$
\mathbb{V}\textrm{ar}[\pi_L^{ML}(\varphi)] = \Bigg(\frac{\mathbb{V}\textrm{ar}[\varphi(X^{1,0})]}{N_0}+\sum_{l=1}^L\frac{
\mathbb{V}\textrm{ar}[\varphi(X^{1,l})-\varphi(\tilde{X}^{1,l-1})]
}{N_l}\Bigg).
$$
Thus if {$\mathbb{V}\textrm{ar}[\varphi(X^{1,l})-\varphi(\tilde{X}^{1,l-1})]$ falls sufficiently fast with $l$, and given an appropriate characterization of the bias $
[\pi_L-\pi](\varphi)$,  and the cost as a function of $l$ it is possible to choose $N_l$ and $L$ to improve upon the i.i.d.~estimator
$$
\frac{1}{N}\sum_{i=1}^N \varphi(X^{i,L}),
$$
where $X^{i,L}$ are i.i.d.~from $\pi_L$.

\subsection{Multilevel Ensemble Kalman--Bucy Filter}

As we have introduced the underlying concepts of MLMC, we now combine this with the EnKBF to result in the MLEnKBF. The discussion
here was first presented in \cite{CJY20}. As we are considering a ML setting for the EnKBF we require two couplings,
 therefore we run the coupled particle systems of firstly, what we refer to as the multilevel VEnKBF (MLVEnKBF)
\begin{align}
\textbf{(F1)}
\begin{cases}
\xi_{(k+1)\Delta_l}^{i,l} & = \xi_{k\Delta_l}^{i,l} + A\xi_{k\Delta_l}^{i,l}\Delta_l + R_1^{1/2} [\overline{W}_{(k+1)\Delta_l}^i-\overline{W}_{k\Delta_l}^i]  
+ U_{k\Delta_l}^{N,l}\Big([Y_{(k+1)\Delta_l}-Y_{k\Delta_l}] 
\\ &-\Big[C\xi_{k\Delta_l}^{i,l}\Delta_l + R_2^{1/2}[\overline{V}_{(k+1)\Delta_l}^i-\overline{V}_{k\Delta_l}^i]\Big]\Big),\\
\xi_{(k+1)\Delta_{l-1}}^{i,l-1} & =  \xi_{k\Delta_{l-1}}^{i,l-1} + A\xi_{k\Delta_{l-1}}^{i,l-1}\Delta_{l-1} + R_1^{1/2} [\overline{W}_{(k+1)\Delta_{l-1}}^i-\overline{W}_{k\Delta_{l-1}}^i]+ U_{k\Delta_{l-1}}^{N,l-1}\Big([Y_{(k+1)\Delta_{l-1}}-Y_{k\Delta_{l-1}}]  \\
&-\Big[C\xi_{k\Delta_{l-1}}^{i,l-1}\Delta_{l-1}+ R_2^{1/2}[\overline{V}_{(k+1)\Delta_{l-1}}^i-\overline{V}_{k\Delta_{l-1}}^i]\Big]\Big),
\end{cases}
\end{align}

and the  multilevel DEnKBF (MLDEnKBF)

\begin{align}
\textbf{(F2)}
\begin{cases}
\xi_{(k+1)\Delta_l}^{i,l} & = \xi_{k\Delta_l}^{i,l} + A\xi_{k\Delta_l}^{i,l}\Delta_l + R_1^{1/2} [\overline{W}_{(k+1)\Delta_l}^i-\overline{W}_{k\Delta_l}^i]  
+ U_{k\Delta_l}^{N,l}\Big([Y_{(k+1)\Delta_l}-Y_{k\Delta_l}] 
\\ &-\frac{1}{2}\Big[C\xi_{k\Delta_l}^{i,l}\Delta_l + Cm^{N,l}_{k\Delta_l}\Delta_l\Big]\Big),\\
\xi_{(k+1)\Delta_{l-1}}^{i,l-1} & =  \xi_{k\Delta_{l-1}}^{i,l-1} + A\xi_{k\Delta_{l-1}}^{i,l-1}\Delta_{l-1} + R_1^{1/2} [\overline{W}_{(k+1)\Delta_{l-1}}^i-\overline{W}_{k\Delta_{l-1}}^i]+ U_{k\Delta_{l-1}}^{N,l-1}\Big([Y_{(k+1)\Delta_{l-1}}-Y_{k\Delta_{l-1}}]  \\
 &-\frac{1}{2}\Big[C\xi_{k\Delta_{l-1}}^{i,l-1}\Delta_{l-1} + Cm^{N,l-1}_{k\Delta_{l-1}}\Delta_{l-1}\Big]\Big),\\
\end{cases}
\end{align}

where $U_{k\Delta_s}^{N,s}=P_{k\Delta_s}^{N,s}C^{\top}R_2^{-1}$, $s\in\{l-1,l\}$, and our sample covariances and means are defined as
\begin{align*}
P_{k\Delta_l}^{N,l} & =  \frac{1}{N -1}\sum_{i=1}^N (\xi_{k\Delta_l}^{i,l}-m_{k\Delta_l}^{N,l})(\xi_{k\Delta_l}^{i,l}-m_{k\Delta_l}^{N,l})^{\top} ,\\
m_{k\Delta_l}^{N,l} & =  \frac{1}{N }\sum_{i=1}^N \xi_{k\Delta_l}^{i,l},\\
P_{k\Delta_{l-1}}^{N,l-1} & =  \frac{1}{N -1}\sum_{i=1}^N (\xi_{k\Delta_{l-1}}^{i,l-1}-m_{k\Delta_{l-1}}^{N,l-1})(\xi_{k\Delta_{l-1}}^{i,l-1}-m_{k\Delta_{l-1}}^{N,l-1})^{\top}, \\
m_{k\Delta_{l-1}}^{N,l-1} & =  \frac{1}{N }\sum_{i=1}^N \xi_{k\Delta_{l-1}}^{i,l-1},
\end{align*}
and $\xi_0^{i,l}\stackrel{\textrm{i.i.d.}}{\sim}\mathcal{N}_{d_x}(\mathcal{M}_0,\mathcal{P}_0)$, $\xi_{0}^{i,l-1}=\xi_0^{i,l}$.
Then, one has the approximation of $[\eta_t^l-\eta_t^{l-1}](\varphi)$, $t\in\mathbb{N}_0$, $\varphi:\mathbb{R}^{d_x}\rightarrow\mathbb{R}$, given by
\begin{align}
[\eta_t^{N,l} -\eta_t^{N,l-1}](\varphi) = \frac{1}{N}\sum_{i=1}^N[\varphi(\xi_t^{i,l})-\varphi(\xi_t^{i,l-1})].
    \label{eq:EnKBF_coupled}
\end{align}

Note that estimation of the filter at any time $t\in\{\Delta_{l-1},2\Delta_{l-1},\dots\}$ can easily be done in the same way as above. 
Then from this we are interested in estimating the following
\begin{equation}\label{eq:main_est}
\eta_t^{ML}(\varphi):=\eta_t^{N_0,0}(\varphi) + \sum_{l-1}^L [\eta_t^{N_l,l} -\eta_t^{N_l,l-1}](\varphi).
\end{equation}
A concrete description of the MLEnKBF is presented in  \autoref{alg:MLEnKBF}.
\subsection{Error-to-cost rate of MLEnKBF}
The work of Chada et al. \cite{CJY20} presents a ML analysis for the vanilla variant \textbf{(F1)}, with a slight modification. The 
theory derived is presented for a simpler ideal i.i.d.~coupled particle system for $(i,k)\in\{1,\dots,N\}\times\mathbb{N}_0$:
\begin{align}
\zeta_{(k+1)\Delta_l}^{i,l} & = \zeta_{k\Delta_l}^{i,l} + A\zeta_{k\Delta_l}^{i,l}\Delta_l + R_1^{1/2} [\overline{W}_{(k+1)\Delta_l}^i-\overline{W}_{k\Delta_l}^i]  
+ U_{k\Delta_l}^{l}\Big([Y_{(k+1)\Delta_l}-Y_{k\Delta_l}] \label{eq:iid1}\\ &-\Big[C\zeta_{k\Delta_l}^{i,l}\Delta_l + R_2^{1/2}[\overline{V}_{(k+1)\Delta_l}^i-\overline{V}_{k\Delta_l}^i]\Big]\Big), \nonumber \\
\zeta_{(k+1)\Delta_{l-1}}^{i,l-1} & =  \zeta_{k\Delta_{l-1}}^{i,l-1} + A\zeta_{k\Delta_{l-1}}^{i,l-1}\Delta_{l-1} + R_1^{1/2} [\overline{W}_{(k+1)\Delta_{l-1}}^i-\overline{W}_{k\Delta_{l-1}}^i] 
 \label{eq:iid2}\\&+ U_{k\Delta_{l-1}}^{l-1}\Big([Y_{(k+1)\Delta_{l-1}}-Y_{k\Delta_{l-1}}] -\Big[C\zeta_{k\Delta_{l-1}}^{i,l-1}\Delta_{l-1}+ R_2^{1/2}[\overline{V}_{(k+1)\Delta_{l-1}}^i-\overline{V}_{k\Delta_{l-1}}^i]\Big]\Big),\nonumber
\end{align}
where $U_{k\Delta_s}^s=P_{k\Delta_s}^{s}C^{\top}R_2^{-1}$, $s\in\{l-1,l\}$, and  also $\xi_0^{i,l}=\xi_0^{i,l-1}=\zeta_0^{i,l}=\zeta_0^{i,l-1}\stackrel{\textrm{i.i.d.}}{\sim}\mathcal{N}_{d_x}(\mathcal{M}_0,\mathcal{P}_0)$,
with  $\zeta_{(k+1)\Delta_l}^{i,l}|\mathscr{F}_{(k+1)\Delta_l}\stackrel{\textrm{i.i.d.}}{\sim}\mathcal{N}_{d_x}(m^l_{(k+1)\Delta_l},P^l_{(k+1)\Delta_l})$, where
\begin{align}
m_{(k+1)\Delta_l} & =  m_{k\Delta_l} + Am_{k\Delta_l}\Delta_l + U_{k\Delta_l}\Big(
[Y_{(k+1)\Delta_l}-Y_{k\Delta_l}] -Cm_{k\Delta_l}\Delta_l
\Big),\label{eq:iid_mean_rec}\\
P_{(k+1)\Delta_l} & =  P_{k\Delta_l} + \textrm{Ricc}(P_{k\Delta_l})\Delta_l +  (A-P_{k\Delta_l}S)P_{k\Delta_l}(A^{\top}-SP_{k\Delta_l})\Delta_l^2,\label{eq:iid_cov_rec}
\end{align}
and similarly for level $l-1$.
 Then from this i.i.d. coupled system, we set the measure of interest as, with a slight change of notation
\begin{equation}\label{eq:main_est_iid}
\hat{\eta}_t^{ML}(\varphi):=\hat{\eta}_t^{N_0,0}(\varphi) + \sum_{l-1}^L [\hat{\eta}_t^{N_l,l} -\hat{\eta}_t^{N_l,l-1}](\varphi),
\end{equation}
where $[\hat{\eta}_t^{N_l,l} -\hat{\eta}_t^{N_l,l-1}](\varphi)=\frac{1}{N}\sum_{i=1}^N[\varphi(\zeta_{t}^{i,l})-\varphi(\zeta_{t}^{i,l-1})]$.} The motivation behind doing so, was that the i.i.d. coupled system is simpler to work with,
and through a limit analysis one can show that the $\hat{\eta_t}$ coincides with $\eta_t$ as $N \rightarrow \infty$. In particular it has been shown they coincide with high probability. We leave out these technicalities for this 
work as it is not required, but refer to the reader to \cite{CJY20}.  Now provided this new ML estimator, we recall the main theorem from \cite{CJY20} which is an error-to-cost rate of the MSE w.r.t. to the MLEnKBF.

\begin{algorithm}[h!]
\caption{(\textbf{MLEnKBF}) Multilevel Estimation of the Ensemble Kalman--Bucy Filter}
\label{alg:MLEnKBF}
\begin{enumerate}
\item \textbf{Input:} Target level $L\in\mathbb{N}$, start level $l_*\in \mathbb{N}$ such that $l_*<L$, the number of particles on each level $\{N_l\}_{l=l_*}^L$, the time parameter $T\in\mathbb{N}$ and initial independent ensembles $\Big\{\{\tilde{\xi}_0^{i,l_*}\}_{i=1}^{N_{l_*}}, \cdots, \{\tilde{\xi}_0^{i,L}\}_{i=1}^{N_{L}}\Big\}$.
\item \textbf{Initialize:} Set $l=l_*$. For $(i,k)\in \{1,\cdots,N_l\}\times \{0,\cdots,T\Delta_l^{-1}-1\}$, set $\{\xi_0^{i,l}\}_{i=1}^{N_l} = \{\tilde{\xi}_0^{i,l}\}_{i=1}^{N_l}$ .Then using \eqref{eq:eta}, return $\eta_T^{N_l,l}(\varphi)$.
\item \textbf{Iterate:} For $l \in \{l_*+1,\cdots,L\}$ and $(i,k)\in \{1,\cdots,N_l\} \times \{0,\cdots,T\Delta_l^{-1}-1\}$, set $\{\xi_0^{i,l-1}\}_{i=1}^{N_l}=\{\xi_0^{i,l}\}_{i=1}^{N_l}= \{\tilde{\xi}_0^{i,l}\}_{i=1}^{N_l}$ (the one which corresponds to the case used in Step 2.). Then using \eqref{eq:eta}, return $\eta_T^{N_l,l-1}(\varphi)$ \& $\eta_T^{N_l,l}(\varphi)$.
\item \textbf{Output:} Return the multilevel estimation of the EnKBF:
\begin{align}
\label{eq:MLEnKBF_NC}
\eta_T^{ML}(\varphi) = \eta_T^{N_{l_*},l_*}(\varphi) + \sum_{l=l_*+1}^L \{\eta_T^{N_l,l}(\varphi)-\eta_T^{N_l,l-1}(\varphi)\}.
\end{align}
\end{enumerate}
\end{algorithm}

\begin{theorem}
\label{theo:main_theo}
For any $T\in\mathbb{N}$ fixed and $t\in[0,T-1]$ there exists a $\mathsf{C}<+\infty$ such that for any $(L,N_{0:L})\in\mathbb{N}\times\{2,3,\dots\}^{L+1}$,
$$
\mathbb{E}\left[\left\|[\hat{\eta}_t^{ML}-\eta_t](e)\right\|_2^2\right] \leq \mathsf{C}\left(
\sum_{l=0}^L \frac{\Delta_l}{N_l} + \sum_{l=1}^L\sum_{q=1, q\neq l}^L\frac{\Delta_l\Delta_q}{N_lN_q} + \Delta_L^2
\right).
$$
\end{theorem}

\begin{proof}
The proof of the above theorem for \textbf{(F1)} can be found in \cite{CJY20}.
\end{proof}
The above theorem can be understood as that, in order to attain an MSE of order $\mathcal{O}(\epsilon^2)$, the cost associated to this is
of order $\mathcal{O}(\epsilon^{-2}\log(\epsilon)^2)$. So by considering the MLEnKBF this provides a reduction in cost compared to the EnKBF
which, to attain an MSE of the same magnitude, the associated cost is more expensive, i.e. $\mathcal{O}(\epsilon^{-3})$. This is 
relevant to mention because, as we will see in the numerics, we will use this error-to-cost rate to compare our unbiased estimators and the MLEnKBF.

We emphasis that the result from  \autoref{theo:main_theo} is specific to the vanilla variant of the MLEnKBF, i.e. \textbf{(F1)}.
However in this work we aim to establish such a theorem for \textbf{(F2)}. Therefore we provide a similar analysis, which can be found in the Appendix.
We note that we attain the exact same rates as in  \autoref{theo:main_theo} for the MLDEnKBF, which is for the following modified ideal i.i.d. coupled system
\textcolor{black}{
\begin{align}
\zeta_{(k+1)\Delta_l}^{i,l} & = \zeta_{k\Delta_l}^{i,l} + A\zeta_{k\Delta_l}^{i,l}\Delta_l + R_1^{1/2} [\overline{W}_{(k+1)\Delta_l}^i-\overline{W}_{k\Delta_l}^i]  
 \label{eq:iid_1}\\ & + U_{k\Delta_l}^{l}\Big([Y_{(k+1)\Delta_l}-Y_{k\Delta_l}] -\frac{1}{2}\Big[C\zeta_{k\Delta_{l}}^{i,l}\Delta_{l-1} + Cm^{l}_{k\Delta_{l}}\Delta_{l}\Big]\Big), \nonumber \\
\zeta_{(k+1)\Delta_{l-1}}^{i,l-1} & =  \xi_{k\Delta_{l-1}}^{i,l-1} + A\zeta_{k\Delta_{l-1}}^{i,l-1}\Delta_{l-1} + R_1^{1/2} [\overline{W}_{(k+1)\Delta_{l-1}}^i-\overline{W}_{k\Delta_{l-1}}^i] 
\label{eq:iid_2}\\  &+ U_{k\Delta_{l-1}}^{l-1}\Big([Y_{(k+1)\Delta_{l-1}}-Y_{k\Delta_{l-1}}] 
 -\frac{1}{2}\Big[C\zeta_{k\Delta_{l-1}}^{i,l-1}\Delta_{l-1} + Cm^{l-1}_{k\Delta_{l-1}}\Delta_{l-1}\Big]\Big). \nonumber
\end{align}}


 It is important to remember that  \autoref{theo:main_theo} is based on both the ideal i.i.d. coupled systems. For \textbf{(F2)}
the new coupled system is given by the equations \eqref{eq:iid_1} - \eqref{eq:iid_2}. 

\begin{rem}
As conducted in \cite{CJY20, RCJ21}, the numerical experiments were tested using the originally stated ensemble Kalman--Bucy filters, while the 
theory was derived for the i.i.d. couple particle systems. As stated in those previous works, our reasons for this is that the recursion of the MLEnKBF
make it difficult to derive exact multilevel rates, so instead one can consider a simpler system, and use a limiting argument, which is to say as $N \rightarrow \infty$
these systems behave similarly and coincide. This can be found in [ Prop. 2.1., \cite{CJY20}].
\end{rem}

\section{Unbiased Estimation}
\label{sec:unb}
In this section we provide our highlighting contribution, which is the introduction of an unbiased estimator for the EnKBF. Our methodology 
will be based on the debiasing schemes of Rhee and Glynn \cite{RG15}, where we will introduce two unbiased estimators in the filtering context
of the EnKBF. We will review their ideas before providing an algorithm of the unbiased estimation for the EnKBF. 
\subsection{General problem}
Let us begin this section with a brief review of the underlying concept of the unbiased estimation of Monte Carlo \cite{DM11,RG15}.
Let us recall $\eta$ is the measure of interest and that one can only attain bias estimates, such as through appropriate time-discretizations.
Given a step size is $\Delta_l$, one can assume that, for $\varphi:\mathbb{R}^{d_x} \rightarrow \mathbb{R}$,
$$
\lim_{l \rightarrow \infty}\eta^l(\varphi) = \eta(\varphi),
$$
from this we are interested in producing unbiased estimates of the measure $\eta(\varphi)$ as well as finite variance.
One can achieve this through introducing randomization, as discussed in detail in \cite{RG15,MV18}.
Let us further assume that one can attain a sequence of independent random variables $(\Xi_l)_{l \in \mathbb{Z}^+}$
such that
$$
\mathbb{E}[\Xi_l] = \eta^l(\varphi) - \eta^{l-1}(\varphi),
$$
which can be viewed as some form of a coupling between different levels $l \in \mathbb{Z}^+$ of discretization.
Furthermore let $\mathbb{P}_L$ denote a probability mass function on $\mathbb{Z}^+$, where we can sample $L$
from. Now consider the estimate
\begin{equation}
\label{eq:b}
\widehat{\eta(\varphi)}_1 = \frac{\Xi_L}{\mathbb{P}_L(L)}.
\end{equation}
By Theorem 3. \cite{MV18}, $\widehat{\eta(\varphi)}_1$ is an unbiased and finite variance estimator of $\eta(\varphi)$ if
the following holds
$$
\sum_{l \in \mathbb{Z}^+}\frac{1}{\mathbb{P}_L(l)}\mathbb{E}[\Xi^2_l] < + \infty.
$$ 
We highlight that once we can obtain \eqref{eq:b}, we can construct i.i.d. samples in parallel. This is done so, by computing $L_i$
independently of $\mathbb{P}_L$ for $i \in \{1,\ldots,N\}$. Therefore from this one is able to construct unbiased estimator
with MSE of order $\mathcal{O}(N^{-1})$ as
$$
\widehat{\eta(\varphi)} = \frac{1}{N}\sum^{N}_{i=1} \frac{\Xi_L}{\mathbb{P}_L(L_i)}.
$$

\subsection{Strategy}
We now introduce our strategy to compute $(\Xi_l)_{l \in \mathbb{Z}^+}$ using unbiased estimates 
of both ${\eta}^0_t(\varphi)$ and $[{\eta}^l_t - {\eta}^{l-1}_t](\varphi)$. Let $(N_p)_{p \in \mathbb{N}_0}$
 for $N_p \in \mathbb{Z}^+$ be an increasing sequence of positive integers were $\lim_{p \rightarrow \infty}N_p = \infty$.

By consistency we have that, almost surely,
$$
\lim_{p \rightarrow \infty}\eta^{N_p,0}_t (\varphi)  = \eta^0_t(\varphi), \quad \quad \lim_{p \rightarrow \infty}\{[\eta^{N_p,l}_t - \eta^{N_p,l-1}_t](\varphi)\} = [\eta^l_t - \eta^{l-1}_t](\varphi).
$$
where  $\eta^{N_p,l}_t(\varphi)$ is defined as in \eqref{eq:EnKBF_coupled}.
 Furthermore let $\tilde{\eta}_t^{N_p,0}(\varphi)$ be a Monte Carlo estimate of $\eta^0_t(\varphi)$ of $N_p$ samples,
and similarly  $[\tilde{\eta}^{N_p,l}_t - \tilde{\eta}^{N_p,l-1}_t](\varphi)$ of $[\eta^l_t - \eta^{l-1}_t](\varphi)$. We will introduce the specific estimator in the next pages.

Let us suppose that we have the random variables

\begin{align}
\Xi_{l,p}:=
\begin{cases}
[\tilde{\eta}_t^{N_p,0} - \tilde{\eta}_t^{N_{p-1},0}](\varphi), \quad \quad &\textrm{if} \ l=0 \\
 \Big([\tilde{\eta}_t^{N_p,l} - \tilde{\eta}_t^{N_{p},l-1}](\varphi)-[\tilde{\eta}_t^{N_{p-1},l} - \tilde{\eta}_t^{N_{p-1},l-1}](\varphi)\Big), \quad &\textrm{otherwise},
\end{cases}
\label{eq:telescopint_term}
\end{align}
where $\tilde{\eta}_t^{N_{-1},l}(\varphi)=0$ by convention. Now let us set $\Xi_l = \Xi_{l,P}$ where $P$ is sampled according to $\mathbb{P}_P(p)$. Again from \cite{MV18} the expected value

\begin{align}
\mathbb{E}\left[\frac{\Xi_{l}}{\mathbb{P}_P(P)}\right]:=
\begin{cases}
\tilde{\eta}^0_t(\varphi) &\textrm{if} \ l=0 \\
\tilde{\eta}^l_t(\varphi) - \tilde{\eta}^{l-1}_t(\varphi) &\textrm{otherwise}.
\end{cases}
\end{align}
Then finally for each $l \in \mathbb{Z}^+$, the random variable $\frac{\Xi_l}{\mathbb{P}_P(P)}$ has finite variance, provided that
\begin{align}
\mathbb{E}\left[\left\|\frac{\Xi_{l}}{\mathbb{P}_P(P)}\right\|^2_2\right]= \sum_{p \geq 0}\mathbb{P}_P(p)^{-1}\mathbb{E}\left[\left\|\Xi_{l,p}\right\|^2_2\right] <+ \infty,
\label{eq:variance0}
\end{align}
furthermore, the random variable $\Xi_{L,P}(\mathbb{P}_P(P)\mathbb{P}_L(L))^{-1}$ has finite variance if the following condition holds 
\begin{align}
\mathbb{E}\left[\left \|\frac{\Xi_{L}}{\mathbb{P}_P(P)\mathbb{P}_L(L)}\right\|_2^2\right]= \sum_{l \geq 0}\sum_{p \geq 0}\mathbb{P}_P(p)^{-1}\mathbb{P}_L(l)^{-1}\mathbb{E}\left[\left\|\Xi_{l,p}\right\|^2_2\right] <+ \infty.
\label{eq:variance}
\end{align}

We must consider how one can compute the quantities $\tilde{\eta}^{N_p,0}_t(\varphi)$ and $[\tilde{\eta}^{N_p,l}_t - \tilde{\eta}^{N_p,l-1}_t](\varphi)$, which we discuss below.

Let us first consider $\tilde{\eta}^{0}_t(\varphi)$.  To form our approximation with $N_0$ samples we run the EnKBF with $N_0$ samples.
For the approximation, regarding $N_1$ samples, we run the EnKBF independently of the first EnKBF with $N_1-N_0$ samples, and continue
this process ($N_2-N_1, \ldots, N_p - N_{p-1}$ samples.) for any $p \geq 2$. We now define the following notation
\begin{align*}
\tilde{\eta}^{N_{p},0}_t(\varphi)&:= \textcolor{black}{\sum^p_{q=0}} \Big(\frac{N_q - N_{q-1}}{N_p} \Big)\eta_t^{N_q - N_{q-1},0}(\varphi), \\
\eta^{N_q-N_{q-1},0}_t(\varphi)&:=\frac{1}{N_q - N_{q-1}} \sum^{N_q}_{\textcolor{black}{i=N_{q-1}+1}}\varphi(\xi^{i,0}_t),
\end{align*}
where $x_t^{1,0}, \ldots, x_t^{N_0,0}$ are generated from the first EnKBF, $x_t^{N_0+1,0}, \ldots, x_t^{N_1,0}$ secondly and so on. 

Now we finally consider approximating $[\eta^l_n - \eta^{l-1}_n](\varphi)$, which can be achieved by running two independently generated coupled EnKBFs,
and using the same procedure as described above. We now introduce the notation, for $s \in \{l,l-1\}$ 
\begin{align*}
\tilde{\eta}^{N_{p},s}_t(\varphi)&:=  \textcolor{black}{\sum^p_{q=0}} \Big(\frac{N_q - N_{q-1}}{N_p} \Big)\eta_t^{N_q - N_{q-1},s}(\varphi), \\
\eta^{N_q-N_{q-1},s}_t(\varphi)&:=\frac{1}{N_q - N_{q-1}} \sum^{N_q}_{\textcolor{black}{i=N_{q-1}+1}}\varphi(\xi^{i,s}_t).
\end{align*}

We provide a detailed structure of our unbiased estimator through \autoref{alg:unbiased}. As we will discuss in our numerics,
we will introduce two unbiased estimators for the EnKBF. From our methodology we implement a double randomization scheme,
similar to \cite{RG15}, where we randomize over firstly the level and  secondly to obtain unbiased estimates of the increments.
 These will be referred to as the single-term (ST) estimate and coupled sum (CS) estimators. We have given a description of how the implementation
  changes with these in \autoref{alg:unbiased}. Related to our unbiased estimators, is the important question of how one chooses
  the probability mass functions. This will be discussed in the following subsection.

  \begin{rem}
  With the above discussion on the unbiased estimator, we note that we do not prove that our estimator
  is unbiased and has a finite variance. This is for two reasons, firstly because of the difficulty,
  and secondly the multilevel analysis only derived in \cite{CJY20} holds only the i.i.d. filter not the 
  EnKBF described in \eqref{eq:enkbf}. Therefore we leave this for future work.
  \end{rem}
  
  \subsection{Discussion on Costs}
\label{sec:Discussion on the cost}

Double randomization schemes have already been applied in \cite{JLY20} to the particle filter. In the current work we apply an analogous setting cost-wise. In fact, provided that the probability distributions $\mathbb{P}_L$ and $\mathbb{P}_P$ satisfy condition (\ref{eq:variance}), the error-to-cost rates are obtained in the same way as in \cite{JLY20}. If we set $N_{p}=N_{0} 2^{p}$,  $\Delta_{l}=2^{-l}$, $\mathbb{P}_{P}(p) \propto N_{p}^{-1}(p+1) \log _{2}(p+2)^{2}$ and $\mathbb{P}_{L}(l) \propto \Delta_{l}(l+1) \log _{2}(l+2)^{2}$, and we choose $M=\mathcal{O}(\varepsilon^{-2})$ (with $\varepsilon>0$) so that the variance is $\mathcal{O}(\varepsilon^2)$ then the cost to obtain $M$ samples is $\mathcal{O}(\varepsilon^{-2}|\operatorname{log}(\varepsilon)|^{2+\delta})$ for any $\delta>0$ (\cite{JLY20} page 14). Which can be compared to the error to cost rate of the MLEnKBF (in \cite{CJY20}), that has a cost of $\mathcal{O}(\varepsilon^{-2}|\operatorname{log}(\varepsilon)|^{2})$ to attain a MSE of $\mathcal{O}(\varepsilon^2)$. 

In \autoref{sec:num} we use assumptions that bound the second moment of (\ref{eq:telescopint_term}), such that condition (\ref{eq:variance}) holds for the previously specified probability measures.

\begin{algorithm}[h!]
\caption{Unbiased Estimate of the Ensemble Kalman--Bucy Filter}
\label{alg:unbiased}
\begin{algorithmic}[1]
\item{\textbf{Input:} Two positive probability mass functions $\mathbb{P}_L$ and $\mathbb{P}_P$ on $\mathbb{N}_0$. For $i=1,\cdots,M$, run Step 2. independently:}
\item{\textbf{Iterate:} Sample $(l_i,p_i)\in\mathbb{N}_0^2$ from  $\mathbb{P}_L\otimes\mathbb{P}_P$.}
\begin{itemize}
\item[I.]{If $l_i=0$. For $s\in\{0,\dots,p_i\}$, independently generate an EnKBF with $N_s-N_{s-1}$ samples
where $0=N_{-1}<N_0<\cdots<N_{p_i}$ are given (e.g.~$N_{p_i}=N_0\,2^{p_i}$, $N_0\in \mathbb{N}$ fixed). Return
\begin{align*} \overline{\Xi}_{0,p_i} = \frac{1}{\mathbb{P}_L(0)\mathbb{P}_P(p_i)}\Big(\textcolor{black}{{\tilde{\eta}}}_t^{N_{p_i},\textcolor{black}{0}}(\varphi)-\textcolor{black}{{\tilde{\eta}}}_t^{N_{p_i-1},\textcolor{black}{0}}(\varphi)\Big)= \frac{1}{\mathbb{P}_L(0)\mathbb{P}_P(p_i)}\Xi_{0,p_i}.
\end{align*}

} 
\bigskip
\item[II.]{Otherwise  for $s\in\{0,\dots,p_i\}$, independently generate coupled EnKBFs with $N_s-N_{s-1}$ samples
where $0=N_{-1}<N_0<\cdots<N_{p_i}$ are given (e.g.~$N_{p_i}=N_0\, 2^{p_i}$, $N_0\in \mathbb{N}$ fixed). Return
\begin{align*}
\overline{\Xi}_{l_i,p_i} &= \frac{1}{\mathbb{P}_L(l_i)\mathbb{P}_P(p_i)}\Big\{\Big(\textcolor{black}{{\tilde{\eta}}}_t^{N_{p_i},l_i}(\varphi)-\textcolor{black}{{\tilde{\eta}}}_t^{N_{p_i},l_i-1}(\varphi)\Big)-
\Big(\textcolor{black}{{\tilde{\eta}}}_t^{N_{p_i-1},l_i}(\varphi)-\textcolor{black}{{\tilde{\eta}}}_t^{N_{p_i-1},l_i-1}(\varphi)\Big)
\Big\}\\&= \frac{\Xi_{l_i,p_i} }{\mathbb{P}_L(l_i)\mathbb{P}_P(p_i)},
\end{align*}
such that $\tilde{\eta}_t^{N_{p_i},l_i}$ has been computed by using an equally weighted empirical measure of all $N_{p_i}$ simulated samples, as in Step I. For
the \textit{coupled sum estimator} we return
}

\begin{align*}
\overline{\Xi}_{l_i,p_i} =& \sum^{p_i}_{s=0}\frac{1}{\sum^{\infty}_{q=s}\mathbb{P}_L(l_i)\mathbb{P}_P(q)}\Big\{\Big(\textcolor{black}{{\tilde{\eta}}}_t^{N_{s},l_i}(\varphi)-\textcolor{black}{{\tilde{\eta}}}_t^{N_{s},l_i-1}(\varphi)\Big)-
\Big(\textcolor{black}{{\tilde{\eta}}}_t^{N_{s-1},l_i}(\varphi)-\textcolor{black}{{\tilde{\eta}}}_t^{N_{s-1},l_i-1}(\varphi)\Big)
\Big\}    \\
=& \sum^{p_i}_{s=0}\frac{\Xi_{l_i,s}}{\sum^{\infty}_{q=s}\mathbb{P}_L(l_i)\mathbb{P}_P(q)},    
\end{align*}

\end{itemize} 
\item{ \textbf{Output:} Return the unbiased estimator
\begin{align}
\label{eq:unbiased_log_estimate}
\widehat{\textcolor{black}{{\eta}_t}(\varphi)}=\frac{1}{M}\sum_{i=1}^M \overline{\Xi}_{l^i,p^i}.
\end{align}

}
\end{algorithmic}
\end{algorithm}

\section{Numerical Results}
\label{sec:num}

In this section, we conduct numerical experiments related to our unbiased estimators. A brief discussion of the algorithm's choice of parameters is also presented. 
We test our method on a linear Ornstein--Uhlenbeck (OU) process with varying dimension sizes.
Specifically we compute the estimator with dimension size of $d_x=d_y=2$ and a higher dimensional example of $d_x=d_y=500$.
We also conduct our experiments with the two variants of the EnKBF, which are the VEnKBF and the DEnKBF. For each experiment, we compare our unbiased
estimators to that of the MLEnKBF and its corresponding variants. 
The specific form of the OU process is
$$
dX_t  =  A X_t dt + R_1^{1/2} dW_t,
$$
which represents the signal process. For the parameter choices in our experiments, the drift terms and diffusion coefficients $A, C, R_1, R_2$ are randomly generated in order to gain generality, with some regularity conditions. The initial condition is normally distributed with an initial distribution of
$X_0\sim\mathcal{N}_{d_{x}}(6,I_{d_x})$. Our observation process takes the same form as \eqref{eq:data}, it is not necessary to take it as random, as instead, one can choose fixed values. As mentioned we will have two
dimensional examples, specified as $d_x=d_y=2$ and $d_x=d_y=500$. We choose different final times of $T=\{30,80\}$ for the two dimensional example and $T=80$ for the $d_x=d_y=500$ dimensional example. Recall the cost associated to the MLEnKBF is given as  $\sum_{l=0}^L N_l\Delta_l^{-1}$, as in \autoref{sec:Discussion on the cost}, the time discretization and its the number of particles in terms of its respective levels are $\Delta_l=2^{-l}$ and $N_p=N_02^p$.

\begin{figure}[h!]
\centering
\includegraphics[width=0.48\textwidth]{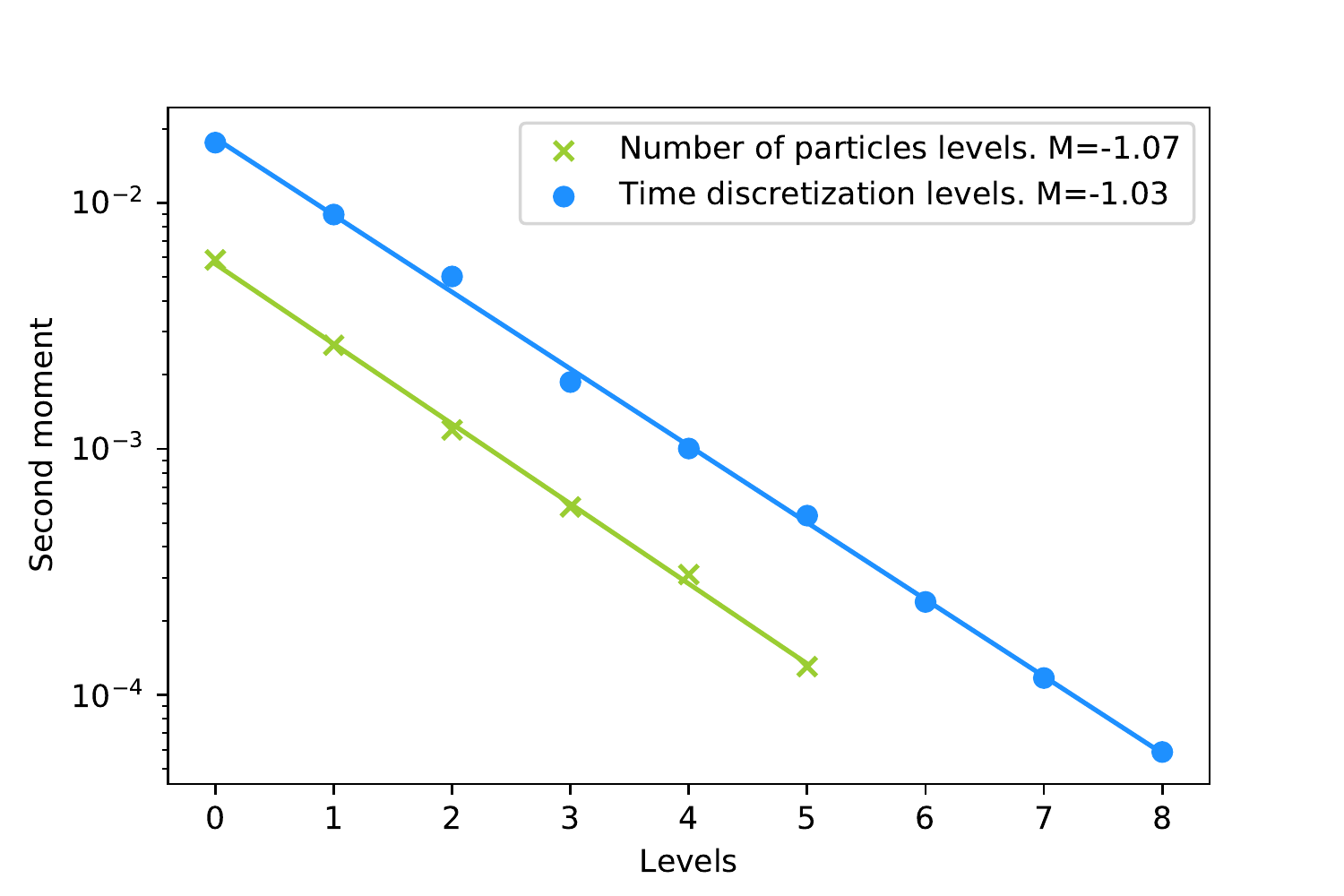}
 \caption{Sampled second moments depending on the time discretization level $l$ (green, $N_0=25$, $p=1$) and on the number of particles level $p$  (blue, $l=2$). }
    \label{fig:second_moment_rate}
\end{figure}

We present an assumption on the second moment of the random variable $\Xi_{L,P}$, where the mechanism that yields the telescoping property hints over
$\mathbb{E}\left[\left(\Xi_{l,p}\right)^2\right]=\mathcal{O}\left(\Delta_l N_p^{-1}\right)$ for $(l,p)\in \mathbb{N}_0^2$, which leaves \eqref{eq:variance} as 
\begin{align*}
\mathbb{E}\left[\left\|\frac{\Xi_{L}}{\mathbb{P}_P(P)\mathbb{P}_L(L)}\right\|^2\right]\leq \mathsf{C}\sum_{l \geq 0}\sum_{p \geq 0}\mathbb{P}_P(p)^{-1}\mathbb{P}_L(l)^{-1}\Delta_l N_p^{-1}.
\label{eq:varianceasumption}
\end{align*}
It is easy to see that the measures considered in \autoref{sec:Discussion on the cost} or the couple of measures  $\mathbb{P}_{P}(p) \propto N_{p}^{-\alpha}$, $\mathbb{P}_{L}(l) \propto \Delta_{l}^{\alpha}$ with $\alpha<1$ leave the variance is finite. In \autoref{fig:second_moment_rate} we present two curves that show the behavior of the second moment when varying either the level $l$ or $p$, where $M$ represents the slope between the logarithms of the levels and the logarithm of the second moment, indicating a direct proportionality relation as in the assumption.

The cost of computing the unbiased estimator is an unbounded random variable related to both measures $\mathbb{P}_L$ and $\mathbb{P}_P$. In most practical cases (see, for example, \cite{JLY20,RG15}), measures that produce finite variance and finite expected cost are difficult or impossible to obtain. In such cases, it is opted to set a measure that leaves the variance finite, sometimes meaning that the expected cost is infinity, but an error-to-cost can be established, even in these cases.

An upper bound for the maximum cost and memory of computing and storing the unbiased estimator is desired, this leads to the truncation of the level measures $\mathbb{P}_{L}$ and $\mathbb{P}_{P}$ in practical implementations of the unbiased estimators. The truncation introduces a bias. Since the maximum level of the truncation can control the bias, the natural choice is to pick measures that yield virtually unbiased estimators, i.e., the choice of the truncation level is such that the bias is comparably small with respect to the target MSE. This way, we keep the primary motivation of the method, obtaining estimators which MSE is given by the variance. 

\begin{figure}[h!]
\centering
\includegraphics[width=0.48\textwidth]{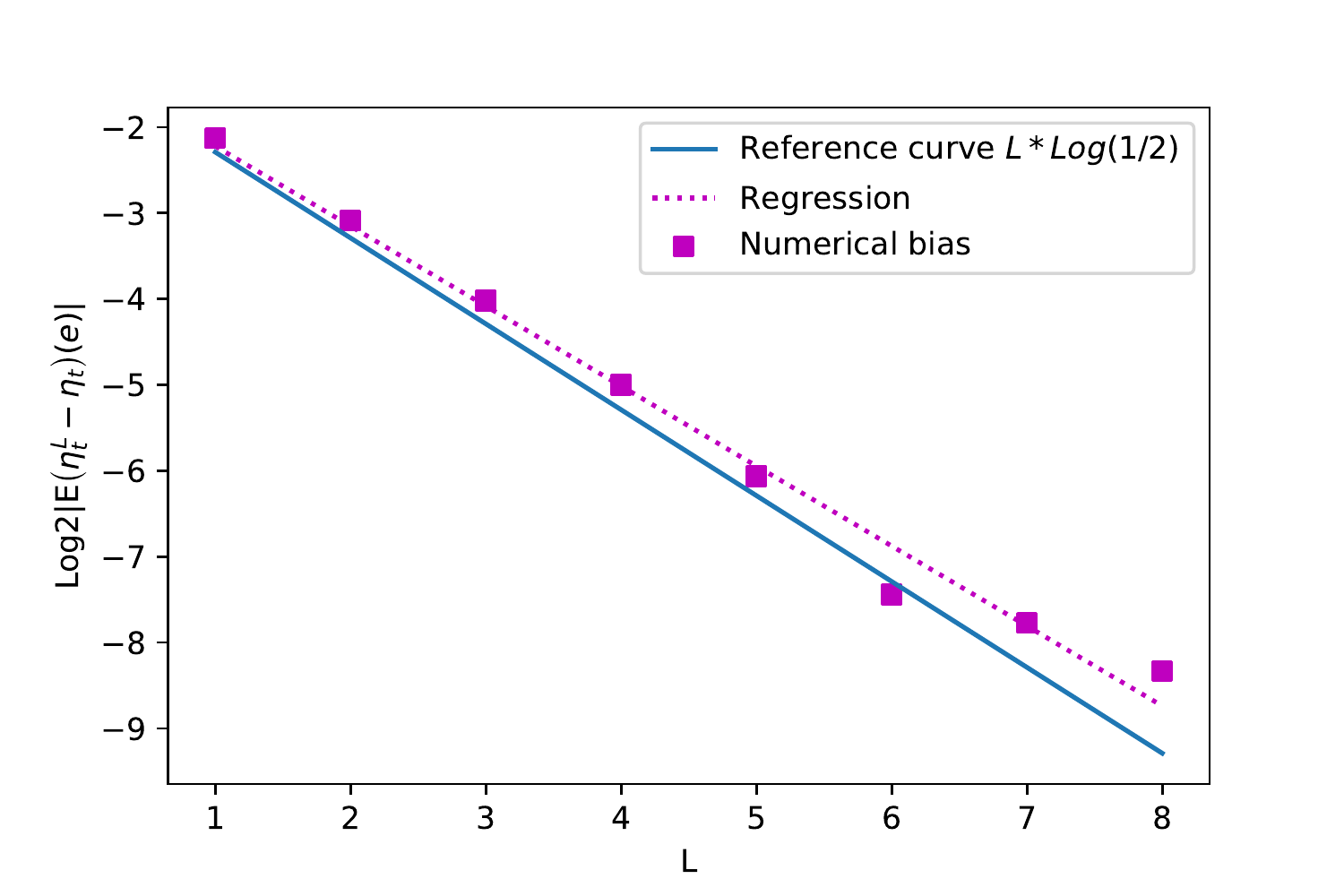}
\includegraphics[width=0.48\textwidth]{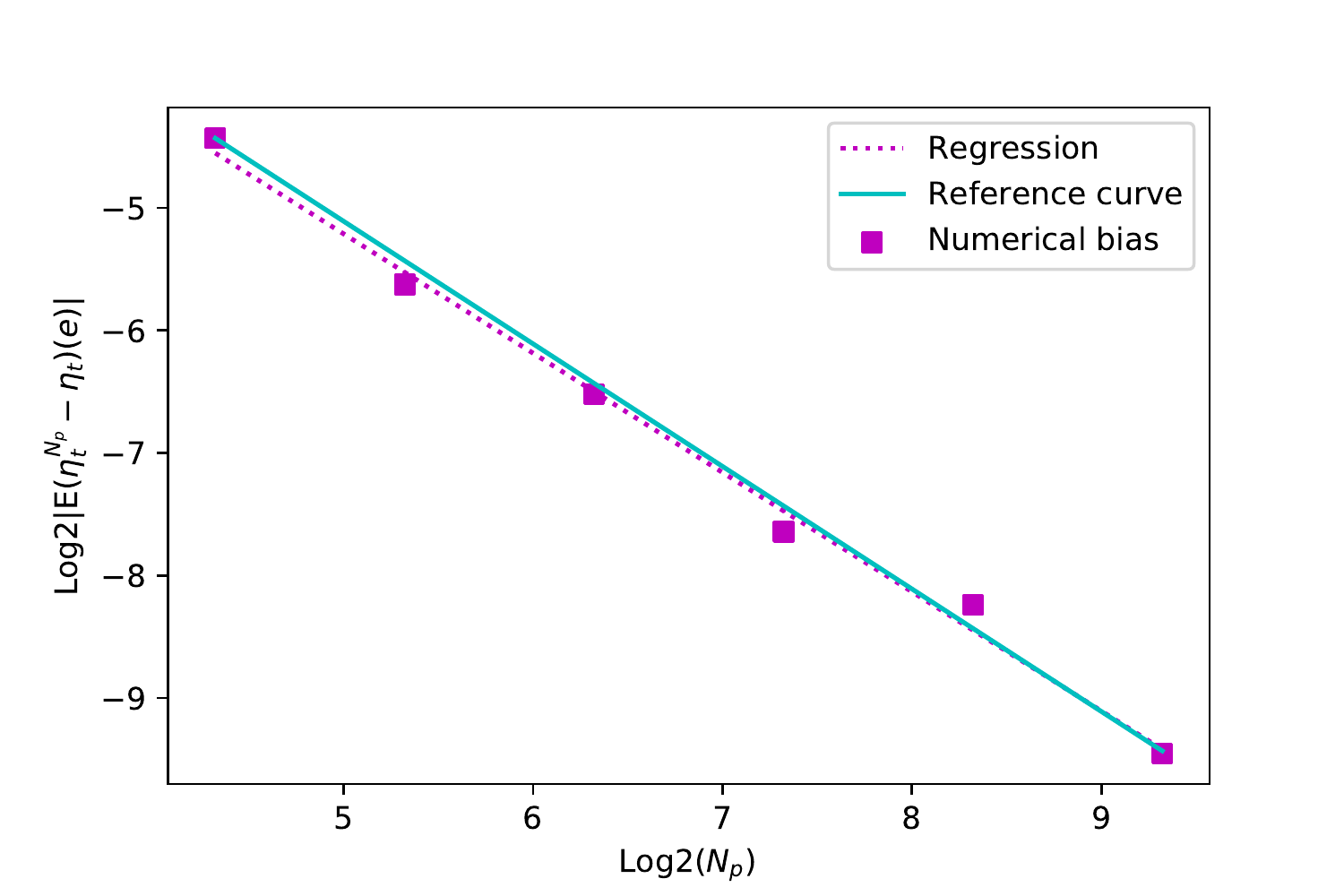}
 \caption{Numerical estimation of the bias depending on the time discretization (left) and on the number of particles (right). We can observe how the numerical bias fits the reference curves, which have slope 1.}
    \label{fig:biases}
\end{figure}

The bias of the estimator can be divided into two sources \begin{align*}
 \left\|\mathbb{E}\left(\eta^{N,l}_t(e)-\eta_t(e)\right)\right\|_2 =&\left\|\mathbb{E}\left(\eta^{N,l}_t(e)-\eta^l_t(e)\right)+
  \eta^{l}_t(e)-\eta_t(e)\right\|_2\\
\leq &  \left\|
  \eta^{l}_t(e)-\eta_t(e)\right\|_2+  \left\|\mathbb{E}\left(\eta^{N,l}_t(e)-\eta^l_t(e)\right)\right\|_2.
\end{align*}
We assume each one of them to be bounded by the following relations 
\begin{align*}
  &   \left\|
  \eta^{l}_t(e)-\eta_t(e)\right\|_2\leq \mathcal{O}(\Delta_l),\\
  &\left\|\mathbb{E}\left(\eta^{N,l}_t(e)-\eta^l_t(e)\right)\right\|_2\leq \mathcal{O}(N_p^{-1}),
\end{align*}
which we tested numerically as seen in \autoref{fig:biases}. In this plot we can observe the relation of the bias depending on the number of levels for both time discretization and the number of particles.

Let $(p_{max} ,l_{max}) \in \mathbb{N}_0\times \mathbb{N}_0$ be the truncation level, $\widetilde{\mathbb{P}}_P(p):\{i\}_{i=0}^{i=p_{max}}\rightarrow (0,1)$, and $\widetilde{\mathbb{P}}_{L}(l):\{i\}_{i=0}^{i=l_{max}}\rightarrow (0,1)$ be the truncated probability measures. The (truncated) unbiased estimator is defined as $\widetilde{\Xi}\equiv \Xi_{\widetilde{L},\widetilde{P}}$, with $\widetilde{L}\sim  \widetilde{\mathbb{P}}_L$ and $\widetilde{P}\sim  \widetilde{\mathbb{P}}_P$. Its computation follows exactly \autoref{alg:unbiased} replacing the measures  $\mathbb{P}_P(p)$ and $\mathbb{P}_{L}(l)$ by $\widetilde{\mathbb{P}}_P(p)$ and $\widetilde{\mathbb{P}}_{L}(l)$, respectively.  For the level of discretization, we will use levels $l=\{3,4,5,6,7\}$. For the number of particles levels we use  $p=\{0,1,2,3,4,5\}$ with $N_0=50$ for the VEnKBF, and $N_0=25$ for the DEnKBF. As stated in \cite{CJY20}
our reason for not choosing smaller levels is that low levels of discretization of the EnKBF can result in potential blow-ups or unstable solutions. Therefore to avoid this, we 
use higher levels. For these same reasons, we use ten dimension block matrices.
It is known that the unbiased estimators and MLMC schemes are closely related \cite{MV18}. This similarity is more recognizable in the truncated setting, given that we control the bias and the variance. This similitude influences the choice of the measures. For the numerical experiments below, we use geometrical probability measures $\mathbb{P}_{P}(p) \propto N_{p}^{-\alpha}$ for $p \in \mathbb{N}_0$,
 and $\mathbb{P}_{L}(l) \propto \Delta_{l}^{\alpha}$ for $l \in \mathbb{N}_0$ with $\alpha<1$ that leave the variance finite. 

\begin{figure}[h!]
\centering
\includegraphics[width=0.48\textwidth]{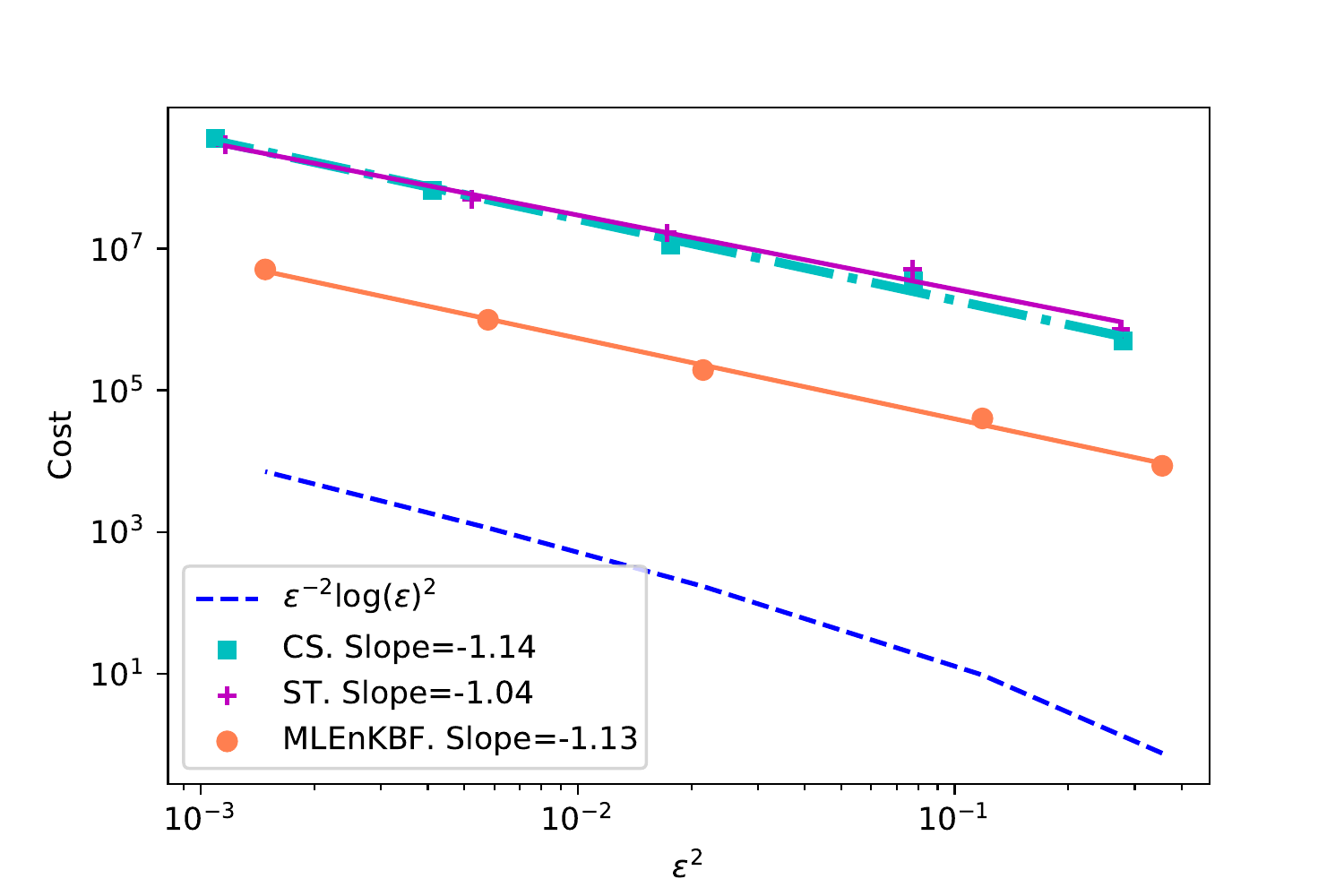}
\includegraphics[width=0.48\textwidth]{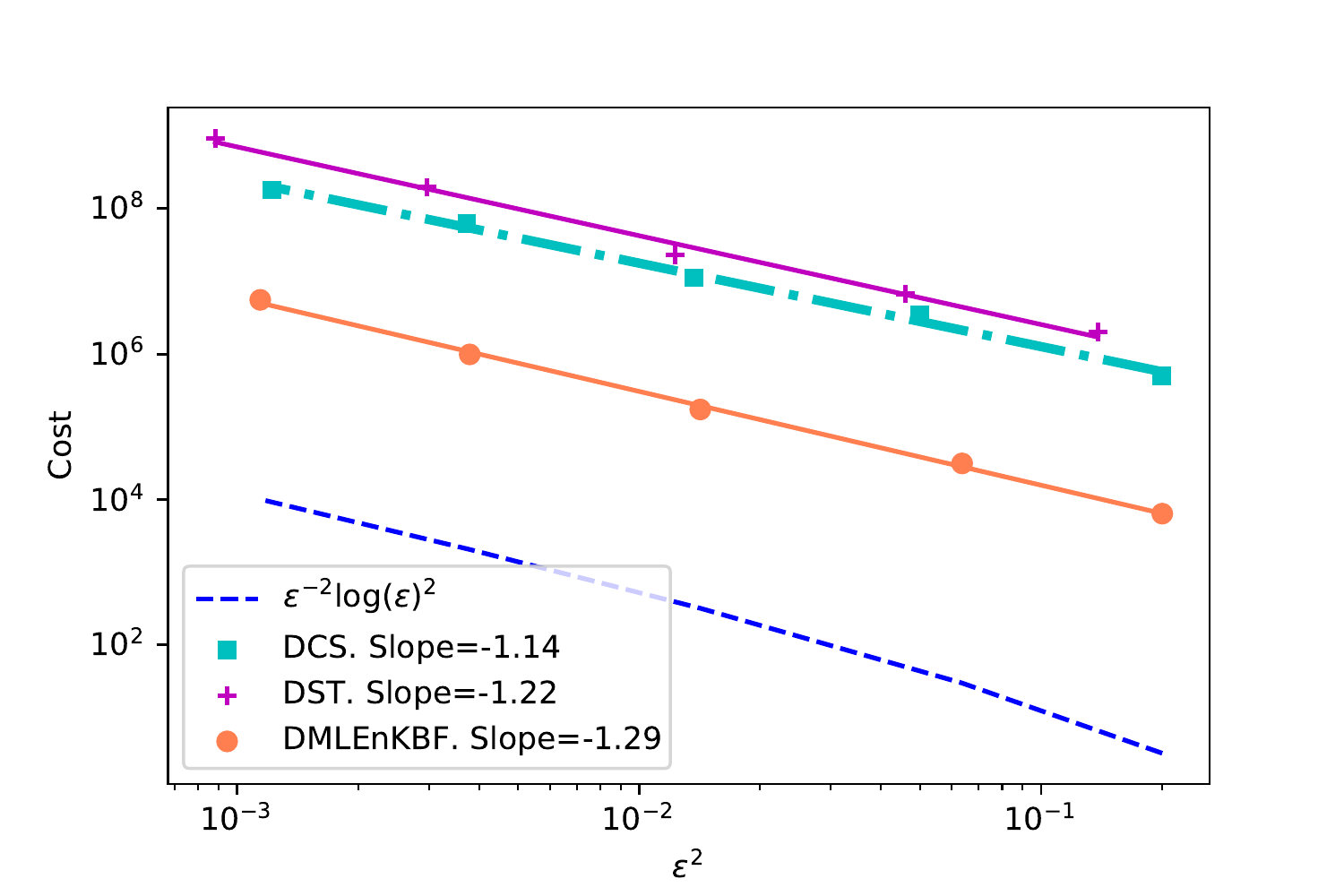}
 \caption{Error (MSE=$\varepsilon^2$) vs cost plots for \textbf{(F1)} (left) and \textbf{(F2)} (right), for the various unbiased estimators,
 and MLEnKBF. Tested for dimension problem of $d_x=d_y=2$ and $T=30$.}
    \label{fig:initial1}
\end{figure}

\begin{figure}[h!]
\centering
\includegraphics[width=0.48\textwidth]{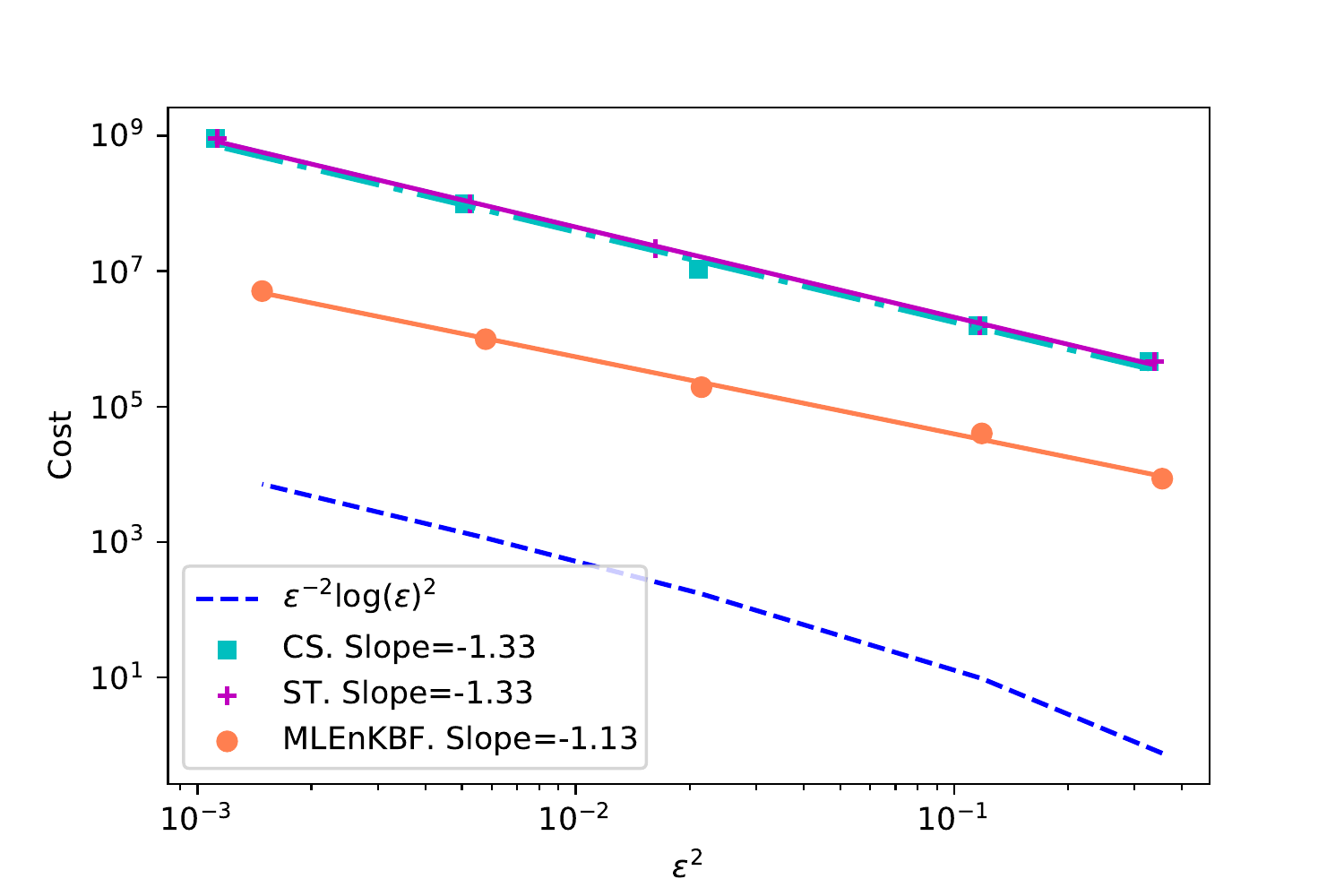}
\includegraphics[width=0.48\textwidth]{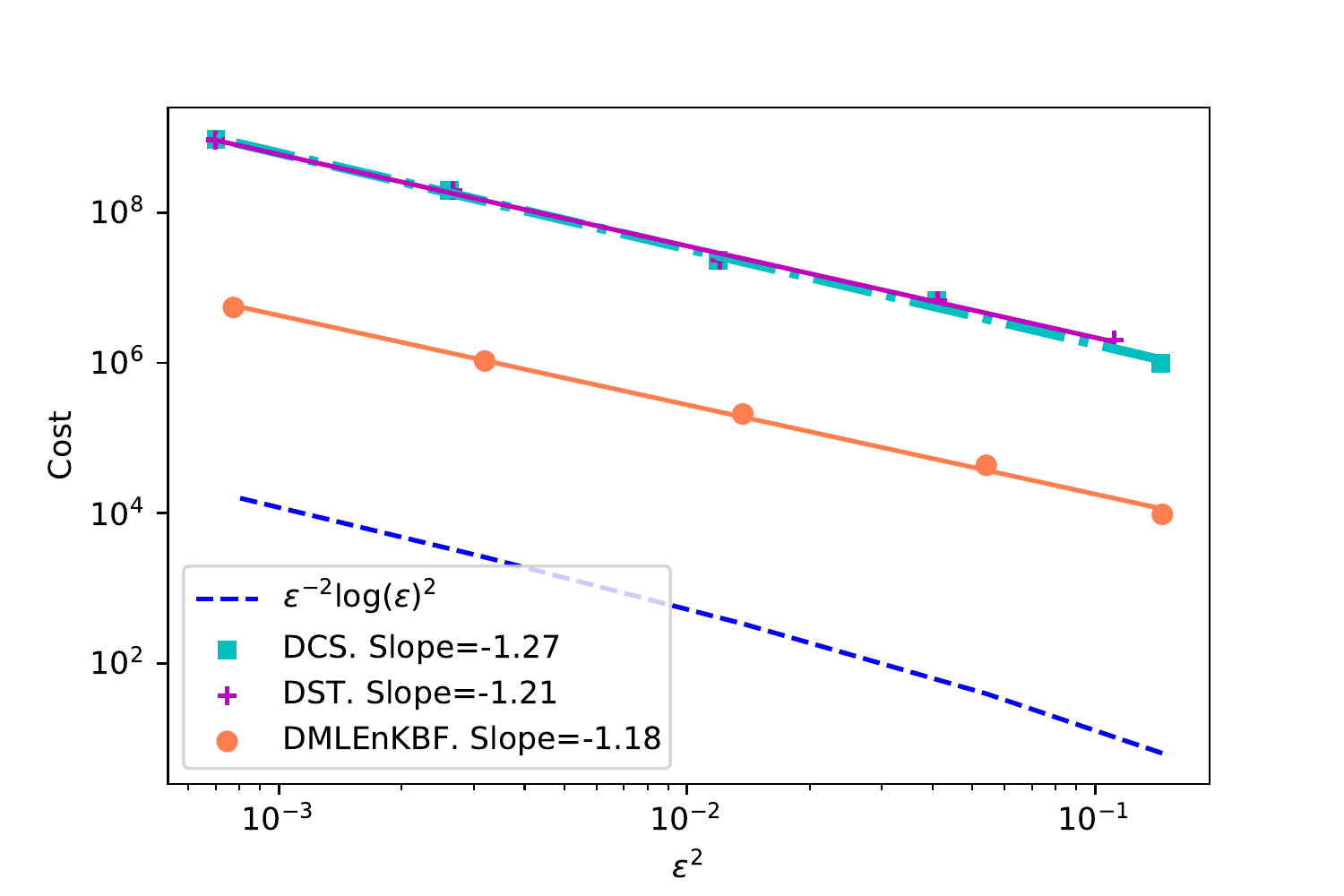}
 \caption{Error vs cost plots for \textbf{(F1)} (left) and \textbf{(F2)} (right), for the various unbiased estimators,
 and MLEnKBF. Tested for dimension problem of $d_x=d_y=2$ and $T=80$.}
    \label{fig:initial2}
\end{figure}

\begin{figure}[h!]
\centering
\includegraphics[width=0.48\textwidth]{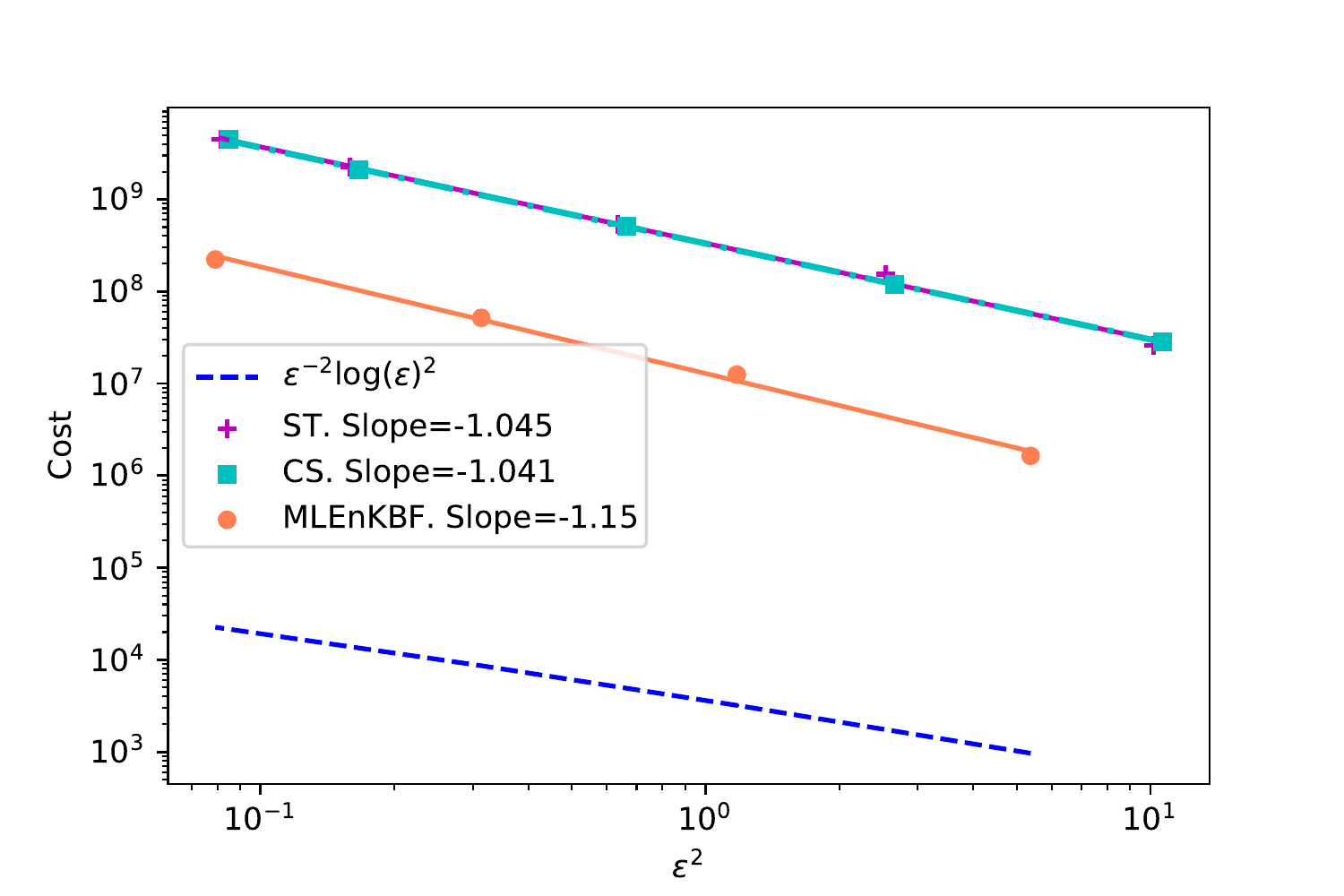}
\includegraphics[width=0.48\textwidth]{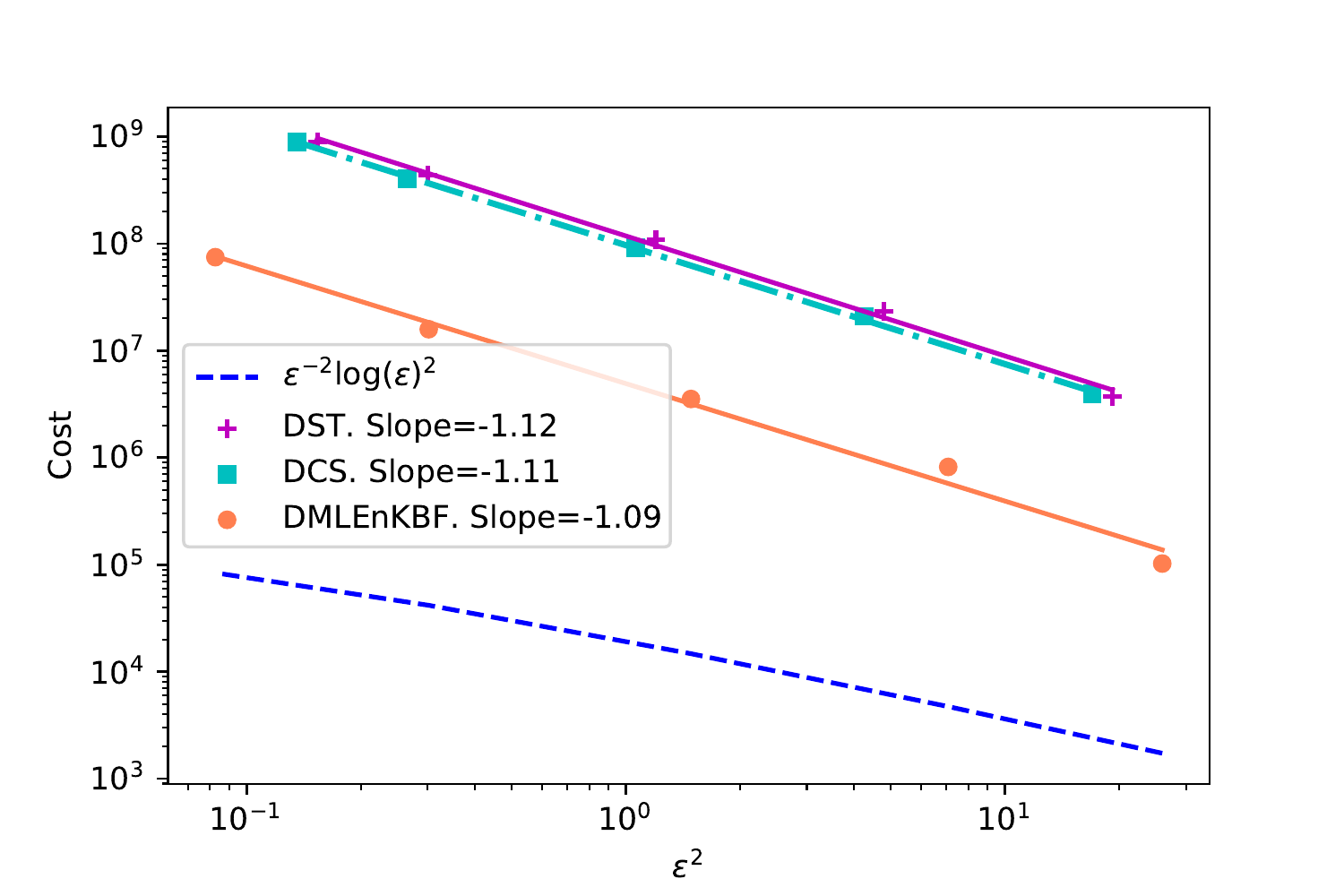}
 \caption{Error vs cost plots for \textbf{(F1)} (left) and \textbf{(F2)} (right), for the various unbiased estimators,
 and MLEnKBF. Tested for dimension problem of $d_x=d_y=500$ and $T=80$.}
    \label{fig:initial3}
\end{figure}

Our primary results are presented in \autoref{fig:initial1} - \autoref{fig:initial3}. Within each of these figures, DCS denotes deterministic coupled sum, DST as deterministic single term, CS denotes coupled sum and ST denotes single term. As we firstly observe from \autoref{fig:initial1}, the MSE-to-cost rates attained for the unbiased schemes match that of both MLEnKBF variants. As one would expect, the cost to attain the same MSE is lower for the MLEnKBF.  Extending this to  \autoref{fig:initial2}, with a slight difference of $T=80$, we notice no significant change with rates and costs roughly the same. Finally, moving onto \autoref{fig:initial3} we modify the dimension of our problem to $d_x=d_y=500$, where we notice that the cost to attain similar order of MSEs, as in the previous examples, requires more computational effort, hence the lower MSEs. However, in terms of the rates, they are similar to what is attained in the previous examples, which should be between $-1.04$ and $-1.3$.

\newpage

\section{Conclusion}
\label{sec:conc}

The development of unbiased estimators is of high relevance for mathematical \& statistical fields, which require the computation of high-dimensional equations.
 In this work we considered such that, aimed at data assimilation for the ensemble Kalman--Bucy filter, which is known to perform well for high-dimensional problems. 
 We proposed two new unbiased estimators, based on the work of Rhee et al. \cite{GR14,RG15} which exploits the use of randomization to produce unbiasedness. 
 We tested our estimators on a relatively high-dimensional Ornstein-Uhlenbeck  process where we compared it to the multilevel EnKBF. As our results suggest, 
 the complexity rate of the MSE-to-error matches what was proved in \cite{CJY20}, and also this  work. For future work, as mentioned earlier, it would be of interest 
 to see if one could attain theoretical justifications for the unbiased estimators, related to proving unbiasedness and finite variance. Another direction is to see how 
 such rates can be applied for nonlinear problems, which is a challenge in itself, and would also be required for the MLEnKBF presented.  One could also aim to
 exploit techniques such as localization or variance inflation \cite{WT19}. These techniques are known to aid, for systems with low
 ensemble sizes, and thus could be implemented in conjunction with the multilevel and unbiased EnKBFs. A final potential direction is the inclusion of such unbiased techniques for the EnKF applied to inverse problems, which can improve the MSE-to-cost as this avenue has not been explored \cite{CST19,ILS13}.

\section*{Acknowledgments}
This work was supported by KAUST baseline funding.

\appendix

\section{Proofs}\label{sec:app_proofs}

\textcolor{black}{In this Appendix we extend the results of \cite{CJY20} to the setting of the MLDEnKBF, where the modification arises from \textbf{(F2)}. Much of the analysis follows almost directly,
however some the results require careful modifications. Here we present the results which are required. Our first result which is required is are the mean and covariance recursion formulas, which
is presented in the following proposition.}

\begin{prop}\label{prop:denkf_rec}
For any $(k,l,N)\in\mathbb{N}_0\times\mathbb{N}_0\times\{2,3,\dots\}$ we have:\\\\
(i) Mean Recursion:
\begin{equation}\label{eq:enkf_mean_rec}
m_{(k+1)\Delta_l}^N - m_{k\Delta_l}^N   =  \Big(A - \textcolor{black}{P_{k\Delta_l}^NS}\Big)\Delta_l m_{k\Delta_l}^N + U_{k\Delta_l}^N[Y_{(k+1)\Delta_l}-Y_{k\Delta_l}]) + \alpha_{k\Delta_l}^N\frac{Z_k}{\sqrt{N}},
\end{equation}
or alternatively
\begin{equation}\label{eq:enkf_mean_rec2}
m^N_{(k+1)\Delta_l}= B^N_{k\Delta_l}m_{k\Delta_l}^N + U_{k\Delta_l}^N\Big([Y_{(k+1)\Delta_l}-Y_{k\Delta_l}]\textcolor{black}{- \frac{C}{2}m^N_{k\Delta_l} \Delta_l}\Big) + \alpha_{k\Delta_l}^N\frac{Z_k}{\sqrt{N}}.
\end{equation}
where $Z_k = \frac{1}{\sqrt{N}}\sum_{i=1}^N \omega_k^i$, independently of all other random variables for each $(i,k)\in\{1,\dots,N\}\times \mathbb{N}_0$ $\omega_k^i \stackrel{\textrm{i.i.d.}}{\sim}\mathcal{N}_{d_x}(0,I)$ and $\textcolor{black}{\alpha_{k\Delta_l}^N=R_1^{1/2}\Delta_l^{1/2}}$. \\\\
(ii) Covariance Recursion:
\begin{align}
\label{eq:enkf_cov_rec}
P_{(k+1)\Delta_l}^N - P_{k\Delta_l}^N & =\textrm{\emph{Ricc}}(P_{k\Delta_l}^N)\Delta_l + \textrm{\emph{SRicc}}(P_{k\Delta_l}^N)\Delta_l^2
\\& \nonumber +\alpha_{k\Delta_l}^N \Big(\tfrac{1}{N-1}\sum_{i=1}^N(\omega_k^i-\bar{\omega}_k)(\omega_k^i-\bar{\omega}_k)^{\top} - I \Big)\alpha_{k\Delta_l}^N \\ \nonumber &+2\textrm{\emph{Sym}}\Big(\alpha_{k\Delta_l}^N\Big(\tfrac{1}{N-1}\sum_{i=1}^N(\omega_k^i-\bar{\omega}_k)(\xi_{k\Delta_l}^i-m_{k\Delta_l}^N)^{\top}\Big)(B_{k\Delta_l}^N)^{\top}\Big),  \nonumber
\end{align}
where $\bar{\omega}_k=\tfrac{1}{N}\sum_{i=1}^N \omega_k^i$ and \textcolor{black}{$\mathrm{SRicc}(P) =(A-\frac{PS}{2})P(A^{\top}-\frac{SP}{2})$}.

\end{prop}

\begin{proof}
Using the notation of the statement of the result, for the d-EnKF, the recursion \eqref{eq:enkf_ps} becomes, for $i\in\{1,\dots,N\}$:
\begin{equation}\label{eq:enkf_mod_rep}
\xi_{(k+1)\Delta_l}^i = B_{k\Delta_l}^N \xi_{k\Delta_l}^i + U_{k\Delta_l}^N\Big([Y_{(k+1)\Delta_l}-Y_{k\Delta_l}] \textcolor{black}{- \frac{C}{2}m^N_{k\Delta_l} \Delta_l}\Big)+ \alpha_{k\Delta_l}^N \omega_k^i.
\end{equation}

To establish \eqref{eq:enkf_mean_rec}, we have by using \eqref{eq:enkf_mod_rep} that
\begin{align*}
m_{(k+1)\Delta_l}^N - m_{k\Delta_l}^N   & =  \frac{1}{N}\sum_{i=1}^N\Big\{\xi_{(k+1)\Delta_l}^i-\xi_{k\Delta_l}^i\Big\} \\
& =  \frac{1}{N}\sum_{i=1}^N\Big\{(B_{k\Delta_l}^N -I)\xi_{k\Delta_l}^i + U_{k\Delta_l}^N\Big([Y_{(k+1)\Delta_l}-Y_{k\Delta_l}] \textcolor{black}{- \frac{C}{2}m^N_{k\Delta_l} \Delta_l}\Big) + \alpha_{k\Delta_l}^N \omega_k^i\Big\}\\
& =  (B_{k\Delta_l}^N -I)\frac{1}{N}\sum_{i=1}^N\xi_{k\Delta_l}^i + U_{k\Delta_l}^N\Big([Y_{(k+1)\Delta_l}-Y_{k\Delta_l}] \textcolor{black}{- \frac{C}{2}m^N_{k\Delta_l} \Delta_l}\Big) + \alpha_{k\Delta_l}^N\frac{1}{N}\sum_{i=1}^N\omega_k^i\\
& =  \Big(A - \textcolor{black}{\frac{P_{k\Delta_l}^NS}{2} - \frac{P_{k\Delta_l}^NC^{\top}R_2^{-1}C}{2}}\Big)\Delta_l m_{k\Delta_l}^N + U_{k\Delta_l}^N [Y_{(k+1)\Delta_l}-Y_{k\Delta_l}] + \alpha_{k\Delta_l}^N\frac{Z_k}{\sqrt{N}} \\
&= \Big(A - \textcolor{black}{{P_{k\Delta_l}^NS}}\Big)\Delta_l m_{k\Delta_l}^N + U_{k\Delta_l}^N[Y_{(k+1)\Delta_l}-Y_{k\Delta_l}] + \alpha_{k\Delta_l}^N\frac{Z_k}{\sqrt{N}}.
\end{align*}

\textcolor{black}{For \eqref{eq:enkf_cov_rec} we have
\begin{equation}
\label{eq:diff_p1}
P_{(k+1)\Delta_l}^N - P_{k\Delta_l}^N = \frac{1}{N-1}\sum_{i=1}^N (\tilde{\xi}_{(k+1)\Delta_l}^i)(\tilde{\xi}_{(k+1)\Delta_l}^i)^{\top}- P_{k\Delta_l}^N,
\end{equation}
where we have used $\tilde{\xi}_{(k+1)\Delta_l}^i := \xi_{(k+1)\Delta_l}^i - m_{(k+1)\Delta_l}^N$. Expanding this, and using \eqref{eq:enkf_mean_rec2} 
and \eqref{eq:enkf_mod_rep}, we have
\begin{align}
\nonumber
\tilde{\xi}_{(k+1)\Delta_l}^i &:= \xi_{(k+1)\Delta_l}^i - m_{(k+1)\Delta_l}^N \\
\nonumber
&=B^N_{k\Delta_l}\xi^i_{k\Delta_l} + U_{k\Delta_l}^N\Big([Y_{(k+1)\Delta_l}-Y_{k\Delta_l}] \textcolor{black}{- \frac{C}{2}m^N_{k\Delta_l} \Delta_l}\Big) + \alpha^N_{k\Delta_l}\omega^i_k \\
\nonumber
&- B_{k\Delta_l}^N m_{k\Delta_l}^N - U_{k\Delta_l}^N\Big([Y_{(k+1)\Delta_l}-Y_{k\Delta_l}] \textcolor{black}{- \frac{C}{2}m^N_{k\Delta_l} \Delta_l}\Big) - \alpha_{k\Delta_l}^N\frac{Z_k}{\sqrt{N}}  \\
\label{eq:diff_p2}
&= B^N_{k\Delta_l}(\xi^i_{k\Delta_l} - m^N_{k\Delta_l}) + \alpha^N_{k\Delta_l}\Big(\omega^i_k - \frac{1}{N}\sum^N_{i=1}\omega^i_k\Big).
\end{align}
Then  by defining 
$\tilde{\omega}^i_{k} = \omega^i_k - \frac{1}{N}\sum^N_{i=1}\omega^i_k$, and substituting \eqref{eq:diff_p2} in \eqref{eq:diff_p1}, we have
\begin{align}
\nonumber
P_{(k+1)\Delta_l}^N - P_{k\Delta_l}^N &= \frac{1}{N-1}\sum^N_{i=1}(B^N_{k\Delta_l}\tilde{\xi}_{(k)\Delta_l}^i + \alpha^N_{k\Delta_l}\tilde{\omega}^i_k)(B^N_{k\Delta_l}\tilde{\xi}_{(k)\Delta_l}^i + \alpha^N_{k\Delta_l}\tilde{\omega}^i_k)^{\top} - P_{k\Delta_l}^N \\
\label{eq:diff_p3}
&= \frac{1}{N-1}\sum^N_{i=1}\Big(B^N_{k\Delta_l}\tilde{\xi}^i_{k\Delta_l}(\tilde{\xi}^i_{k\Delta_l})^{\top}(B^{N}_{k\Delta_l})^{\top} + B^N_{k\Delta_l}\tilde{\xi}^i_{k\Delta_l}(\alpha^N_{k\Delta_l})^{\top}(\tilde{\omega}^{i}_{k})^{\top} \\
\nonumber
&+ \alpha^N_{k\Delta_l}\tilde{\omega}^i_k (\tilde{\xi}^i_{k\Delta_l})^{\top}(B^N_{k\Delta_l})^{\top} +  \alpha^N_{k\Delta_l}\tilde{\omega}^i_k(\tilde{\omega}^i_k)^{\top}( \alpha^N_{k\Delta_l})^{\top}\Big) - 
P_{k\Delta_l}^N.
\end{align}
For the the first and last term of \eqref{eq:diff_p3} we can express it as
\begin{align} \nonumber
 &B_{k\Delta_l}^NP_{k\Delta_l}^N(B_{k\Delta_l}^N)^{\top} - P^N_{k\Delta_l} \\
& =  \Bigg(\Big(P_{k\Delta_l}^N-AP_{k\Delta_l}^N\Delta_l - \frac{P_{k\Delta_l}^NSP_{k\Delta_l}^N\Delta_l}{2}\Big)\Big(I+A^{\top}\Delta_l - \frac{SP_{k\Delta_l}^N\Delta_l}{2}\Big)\Bigg) - P^N_{k\Delta_l}\nonumber\\
& =  P_{k\Delta_l}^NA^{\top}\Delta_l - P_{k\Delta_l}^NSP_{k\Delta_l}^N\Delta_l + A P_{k\Delta_l}^N\Delta_l ++ \textcolor{black}{\Delta_l^2\Big(A-\frac{P^N_{k\Delta_l}S}{2}\Big)P^N_{k\Delta_l}\Big(A^{\top}-\frac{SP^N_{k\Delta_l}}{2}\Big)}\nonumber\\
& =   \textrm{Ricc}(P_{k\Delta_l}^N)\Delta_l  + \textcolor{black}{\Delta_l^2\Big(A-\frac{P^N_{k\Delta_l}S}{2}\Big)P^N_{k\Delta_l}\Big(A^{\top}-\frac{SP^N_{k\Delta_l}}{2}\Big)  - (\alpha_{k\Delta_l}^N)^2}, \label{eq:term_p}
\end{align}
where we know that $(\alpha_{k\Delta_l}^N)^2 = (R^{1/2}\Delta_l^{1/2})^2$. Thus substituting \eqref{eq:term_p} in \eqref{eq:diff_p3} results in \eqref{eq:enkf_cov_rec}, concluding the proof.
}
\end{proof}

\textcolor{black}{Our next results is an $L^q$ bound on the DEnKBF, which is presented in the following lemma.}
{
\begin{lem}\label{lem:xi_lq}
For any $(q,k,l)\in(0,\infty)\times\mathbb{N}_0^2$ there exists a $\mathsf{C}<+\infty$ such that for any $N \geq 2$ and  $i\in \{1,\dots,N\}$:
$$
\mathbb{E}[|\xi_{k\Delta_l}^i(j)|^q]^{1/q} \leq \mathsf{C}.
$$
\end{lem}
}

\begin{proof}
The proof consists of deriving a recursion on $k$ noting that the case $k=0$ is trivial as $\xi^i_0(j)\stackrel{\textrm{i.i.d.}}{\sim}\mathcal{N}(\mathcal{M}_0(j),\mathcal{P}_0(j,j))$, $i\in\{1,\dots,N\}$.

We begin with the following upper-bound, that can be derived by combining \eqref{eq:enkf_mod_rep} along with the Minkowski inequality,
$$
\mathbb{E}[|\xi_{k\Delta_l}^i(j)|^q]^{1/q} \leq \sum_{j=1}^3 T_j,
$$
where
\begin{align*}
T_1 & =  \mathbb{E}\Big[\Big|\sum_{j_1=1}^{d_x} \textcolor{black}{B^N_{k \Delta_l}}\xi_{k\Delta_l}^{i}(j_1)\Big|^q\Big]^{1/q},\\
T_2 & =  \mathbb{E}\Big[\Big|\sum_{j_1=1}^{d_x}\sum_{j_2=1}^{d_x}P_{k\Delta_l}^N(j,j_1)\textcolor{black}{C^{\top} R_2^{-1}(j_1,j_2)}[Y_{(k+1)\Delta_l}-Y_{k\Delta_l}](j_2)\Big|^q\Big]^{1/q},\\
T_3 & =  \mathbb{E}\Big[\Big|\sum_{j_1=1}^{d_x}\alpha_{k\Delta_l}^N(j,j_1) \omega_k^i(j_1)\Big|^q\Big]^{1/q}, \\
T_4 & = \textcolor{black}{\mathbb{E}\Big[\Big|\sum_{j_1=1}^{d_x}\sum_{j_2=1}^{d_x}P_{k\Delta_l}^N(j,j_1)\textcolor{black}{C^{\top} R_2^{-1}(j_1,j_2)}\frac{m^N_{k\Delta_l}(j_1)\Delta_l}{2}\Big|^q\Big]^{1/q}}.
\end{align*}
The terms $T_1-T_3$ are derived in the work of \cite{CJY20}, with the slight modification of now aiming to bound $T_4$. For completeness we state these bounds below
\begin{align}
\label{eq:t1}
T_1 &\leq  \mathsf{C}\max_{j_1\in\{1,\dots,d_x\}}\{\mathbb{E}[|\xi_{k\Delta_l}^{i}(j_1)|^{4q}]^{1/(4q)}\},\\
\label{eq:t2}
T_2 &\leq   \mathsf{C}\max_{j_1\in\{1,\dots,d_x\}}\{\mathbb{E}[|\xi_{k\Delta_l}^{i}(j_1)|^{4q}]^{1/(4q)}\},\\
\label{eq:t3}
T_3 &\leq  \mathsf{C}\left(\max_{j_1\in\{1,\dots,d_x\}}\{\mathbb{E}[|\xi_{k\Delta_l}^{i}(j_1)|^{4q}]^{1/(4q)}\}+\Delta_l^{1/2}\right).
\end{align}
For $T_4$ by Minkowski and Cauchy-Schwarz 
\textcolor{black}{
\begin{align}
\nonumber
T_4 &\leq \sum_{j_1=1}^{d_x}\sum_{j_2=1}^{d_x}\Big|{\frac{C^{\top} R_2^{-1}(j_1,j_2)\Delta_l}{2}}\Big|\mathbb{E}[|P_{k\Delta_l}^N(j,j_2)|^{2q}]^{1/(2q)}\mathbb{E}[|m^N_{k\Delta_l}(j_1)|^{2q}]^{1/(2q)}, \\ \nonumber
&\leq \mathsf{C} \Big(\mathbb{E}[|[\xi_{k\Delta_l}^{i}-m_{k\Delta_l}](j_2) [\xi_{k\Delta_l}^{i}-m_{k\Delta_l}](j)|^{2q}]^{1/(2q)}\\
&+\mathbb{E}[|m_{k\Delta_l}^{N}(j_1)|^{2q}]^{1/(2q)}\Big). \label{eq:t4} 
\end{align}}
Using similar arguments as before, and applying Young's and Jensen's inequality results in $T_4 \leq \mathsf{C}$. Therefore by combining all terms \eqref{eq:t1} - \eqref{eq:t4}
we have
\textcolor{black}{
\begin{align*}
\max_{j_1\in\{1,\dots,d_x\}}\{\mathbb{E}[|\xi_{(k+1)\Delta_l}^{i}(j_1)|^{q}]^{1/(q)}\} &\leq   \mathsf{C}\Big(
\max_{j_1\in\{1,\dots,d_x\}}\{\mathbb{E}[|\xi_{0}^{i}(j_1)|^{4q(k+1)}]^{1/(4q(k+1)}\}\\&+(k+1)\Delta_l^{1/2}\Big),
 \end{align*}}
 which completes the proof.
\end{proof}

Given the results from \autoref{prop:denkf_rec} and \autoref{lem:xi_lq}, the following results follow directly as in \cite{CJY20}.

\begin{lem}\label{lem:p_lq}
{For any $(q,k,l)\in(0,\infty)\times\mathbb{N}_0^2$ there exists a $\mathsf{C}<+\infty$ such that for any $(j_1,j_2,N)\in\{1,\dots,d_x\}^2\times\{2,3,\dots\}$:
$$
\mathbb{E}\Big[\Big|P_{k\Delta_l}^N(j_1,j_2)-P_{k\Delta_l}(j_1,j_2)\Big|^q\Big]^{1/q} \leq \frac{\mathsf{C}}{\sqrt{N}}.
$$}
\end{lem}

\begin{theorem}\label{theo:prop}
{
For any $(q,k,l)\in(0,\infty)\times\mathbb{N}_0^2$ there exists a $\mathsf{C}<+\infty$ such that for any $N \geq 2$ and $(j,i)\in\{1,\dots,d_x\}\times \{1,\dots,N\}$:
$$
\mathbb{E}\Big[\Big|\xi_{k\Delta_l}^i(j)-\zeta_{k\Delta_l}^i(j)\Big|^q\Big]^{1/q} \leq \frac{\mathsf{C}}{\sqrt{N}}.
$$}
\end{theorem}

\subsection{Strong error}

\textcolor{black}{Our final result required is the strong rate for the DEnKBF. In order for us to continue to we require a lemma from \cite{CJY20} presented below, which is a bound
on the difference of the covariances. As there is only a constant different the lemma still holds}. 

\begin{lem}\label{lem:p_disc}
For any $T\in\mathbb{N}$ fixed and $t\in[0,T]$ there exists a $\mathsf{C}<+\infty$ such that for any $(l,j_1,j_2)\in\mathbb{N}_0\times\{1,\dots,d_x\}^2$:
$$
\Big|\mathcal{P}_{t}(j_1,j_2)-P_{\tau_t^l}^l(j_1,j_2)\Big| \leq \mathsf{C}\Delta_l.
$$
\end{lem}

\begin{lem}\label{lem:strong_error}
For any $T\in\mathbb{N}$ fixed and $t\in[0,T-1]$ there exists a $\mathsf{C}<+\infty$ such that for any $(l,j,k_1)\in\mathbb{N}_0\times\{1,\dots,d_x\}\times\{0,1,\dots,\Delta_{l}^{-1}\}$:
$$
\mathbb{E}\Big[\Big(\overline{X}_{t+k_1\Delta_{l}}(j)-\overline{X}_{t+k_1\Delta_{l}}^l(j)\Big)^2\Big] \leq \mathsf{C}\Delta_l^2.
$$
\end{lem}

\begin{proof}
By using the $C_2-$inequality four times, one has the upper-bound
$$
\mathbb{E}\Big[\Big(\overline{X}_{t+k_1\Delta_{l}}(j)-\overline{X}_{t+k_1\Delta_{l}}^l(j)\Big)^2\Big] \leq \mathsf{C}\sum_{j=1}^4 T_j,
$$
where
\begin{align*}
T_1 & = \mathbb{E}\Big[\Big(\int_0^{t+k_1\Delta_{l}}\sum_{j_1=1}^{d_x} A(j,j_1)[\overline{X}_{s}(j)-\overline{X}_{\tau_s^l}^l(j)]ds\Big)^2\Big],\\
T_2 & = \mathbb{E}\Big[\Big(\sum_{j_1=1}^{d_x}\sum_{j_2=1}^{d_y}\int_{0}^{t+k_1\Delta_{l}}[\mathcal{P}_{s}(j,j_1)-P_{\tau_s^l}^l(j,j_1)]\tilde{C}(j_1,j_2)dY_s(j_2)
\Big)^2\Big],\\
T_3 &= \mathbb{E}\Big[\Big(\sum_{j_1=1}^{d_x}\sum_{j_2=1}^{d_x}\int_{0}^{t+k_1\Delta_{l}}[\mathcal{P}_{s}(j,j_1)\hat{C}(j_1,j_2)\overline{X}_s(j_2)
-P^l_{\tau_s^l}(j,j_1)\hat{C}(j_1,j_2)\overline{X}_{\tau_s^l}^l(j_2)]ds
\Big)^2\Big],\\
\textcolor{black}{T_4} & = \textcolor{black}{\mathbb{E}\Big[\Big(\sum_{j_1=1}^{d_x}\sum_{j_2=1}^{d_y}\int_{0}^{t+k_1\Delta_{l}}\frac{\tilde{C} C}{2}\Big[ \mathcal{P}_s(j,j_1)[\overline{X}_s(j_2) + \overline{m}_s(j_2)]
 - P^l_{\tau^l_s}(j,j_1)[X^l_{\tau^l_s}(j_2) + \overline{m}^l_{\tau^l_s}(j_2)]\Big]ds\Big)^2\Big]},
\end{align*}
with $\tilde{C}=C^{\top}R_2^{-1/2}$.
We shall focus on upper-bounding each of the terms $T_1,\dots,T_4$ and summing the bounds.

For $T_1$, $d_x-$applications of the $C_2-$inequality along with Jensen's inequality yields
\begin{equation}\label{eq:strong_error1}
T_1 \leq \mathsf{C}\int_0^{t+k_1\Delta_{l}}\max_{j\in\{1,\dots,d_x\}}\mathbb{E}\Big[\Big(\overline{X}_{s}(j)-\overline{X}_{\tau_s^l}^l(j)\Big)^2\Big]ds.
\end{equation}

From \cite{CJY20} For $T_2$, using \eqref{eq:data} we have the following bound
\begin{equation}
\label{eq:strong_error2}
T_2 \leq \mathsf{C} \Delta^2_l.
\end{equation}
Similarly for $T_3$, we can also use the derived bound, which is
\begin{equation}\label{eq:strong_error3}
T_3 \leq \mathsf{C}\Big(\int_0^{t+k_1\Delta_{l}}\max_{j\in\{1,\dots,d_x\}}\mathbb{E}\Big[\Big(\overline{X}_{s}(j)-\overline{X}_{\tau_s^l}^l(j)\Big)^2\Big]ds+\Delta_l^2\Big).
\end{equation}
\textcolor{black}{
For $T_4$ we can simplify this using the $C_2$--inequality
\begin{eqnarray*}
T_4 &=& \mathbb{E}\Big[\Big(\sum_{j_1=1}^{d_x}\sum_{j_2=1}^{d_y}\int_{0}^{t+k_1\Delta_{l}}\frac{\tilde{C} C}{2}\Big[ [\mathcal{P}_s(j,j_1) \overline{X}_s(j_2) -  {P}^l_{\tau^l_s}(j,j_1)\overline{X}^l_{\tau^l_s}(j_2)]
\\&+& \mathcal{P}_s(j,j_1) \overline{m}_s(j_2) -  {P}^l_{\tau^l_s}(j,j_1)\overline{m}^l_{\tau^l_s}(j_2)] \Big] ds\Big)^2 \Big] \\
&\leq& \mathbb{E}\Big[\Big(\sum_{j_1=1}^{d_x}\sum_{j_2=1}^{d_y}\int_{0}^{t+k_1\Delta_{l}}\frac{\tilde{C} C}{2}\Big[ [\mathcal{P}_s(j,j_1) \overline{X}_s(j_2) -  {P}^l_{\tau^l_s}(j,j_1)\overline{X}^l_{\tau^l_s}(j_2)]
\Big]ds\Big)^2\Big] \\
&+& \mathbb{E}\Big[\Big(\sum_{j_1=1}^{d_x}\sum_{j_2=1}^{d_y}\int_{0}^{t+k_1\Delta_{l}}\frac{\tilde{C} C}{2}\Big[ [\mathcal{P}_s(j,j_1) \overline{m}_s(j_2) -  {P}^l_{\tau^l_s}(j,j_1)\overline{m}^l_{\tau^l_s}(j_2)]
\Big]ds\Big)^2\Big] \\
&=:& T_5 + T_6.
\end{eqnarray*}
We can use the following identity for $T_5$
\begin{equation*}
\mathcal{P}_s(j,j_1) \overline{X}_s(j_2) -  {P}^l_{\tau^l_s}(j,j_1)\overline{X}^l_{\tau^l_s}(j_2) = \{\mathcal{P}_s(j,j_1) - {P}^l_{\tau^l_s}(j,j_1) \} \overline{X}_s(j_2)
+ \{ \overline{X}_s(j_2) - \overline{X}^l_{\tau^l_s}(j_2)\}{P}^l_{\tau^l_s}(j,j_1),
\end{equation*}
and similarly done for $T_6$.  \\
Substituting this into $T_5$, and using $d_xd_y$ applications of the $C_2$-- inequality and Jensen's inequality, we have
\begin{align*}
T_5 &=  \mathbb{E}\Big[\Big(\sum_{j_1=1}^{d_x}\sum_{j_2=1}^{d_y} \Big(\int_{0}^{t+k_1\Delta_{l}}\frac{\tilde{C} C}{2} [\{\mathcal{P}_s(j,j_1) - {P}^l_{\tau^l_s}(j,j_1) \} \overline{X}_s(j_2)
+ \{ \overline{X}_s(j_2) - \overline{X}^l_{\tau^l_s}(j_2)\}{P}^l_{\tau^l_s}(j,j_1)] ds\Big)^2\Big] \\
&\leq \sum_{j_1=1}^{d_x}\sum_{j_2=1}^{d_y}\frac{\tilde{C} C}{2}\Big(\int_{0}^{t+k_1\Delta_{l}} \mathbb{E} \Big[[\{\mathcal{P}_s(j,j_1) - {P}^l_{\tau^l_s}(j,j_1)\}\overline{X}_s(j_2)]^2\Big]ds \\
&+  \mathbb{E} \Big[[\{ \overline{X}_s(j_2) - \overline{X}^l_{\tau^l_s}(j_2)\}{P}^l_{\tau^l_s}(j,j_1)]^2\Big]ds\Big) \\
& \leq   \sum_{j_1=1}^{d_x}\sum_{j_2=1}^{d_y}\frac{\tilde{C} C}{2}\Big(\int_{0}^{t+k_1\Delta_{l}} \max_{j \in \{1,\ldots,d_x\}} \Big|\mathcal{P}_s(j,j_1) - {P}^l_{\tau^l_s}(j,j_1)\Big|^2\mathbb{E}[\overline{X}_s(j_2)]^2ds \\
&+  \max_{j \in \{1,\ldots,d_x\}}\mathbb{E} \Big[ \overline{X}_s(j_2) - \overline{X}^l_{\tau^l_s}(j_2)\Big]^2 \max_{j \in \{1,\ldots,d_x\}}|{P}^l_{\tau^l_s}(j,j_1)|^2ds\Big).
\end{align*}
For the first term of $T_5$ we can use \autoref{lem:p_disc}, and for the second term knowing \\ $\max_{j \in \{1,\ldots,d_x\}}|{P}^l_{\tau^l_s}(j,j_1)|^2 \leq \mathsf{C}$ results in the bound as $T_3$, i.e.
\begin{equation}
\label{eq:strong_error4}
T_5 \leq \mathsf{C}\Big(\int_0^{t+k_1\Delta_{l}}\max_{j\in\{1,\dots,d_x\}}\mathbb{E}\Big[\Big(\overline{X}_{s}(j)-\overline{X}_{\tau_s^l}^l(j)\Big)^2\Big]ds+\Delta_l^2\Big).
\end{equation}
Finally for $T_6$, we can apply the same identity where we get an $\mathcal{O}(1)$-term.} 

Therefore combining \eqref{eq:strong_error1}-\eqref{eq:strong_error4} gives the upper-bound
\begin{align*}
\max_{j\in\{1,\dots,d_x\}}\mathbb{E}\Big[\Big(\overline{X}_{t+k_1\Delta_{l}}(j)-\overline{X}_{t+k_1\Delta_{l}}^l(j)\Big)^2\Big] \leq 
\mathsf{C}\Big(\int_0^{t+k_1\Delta_{l}}\max_{j\in\{1,\dots,d_x\}}\mathbb{E}\Big[\Big(\overline{X}_{s}(j)-\overline{X}_{\tau_s^l}^l(j)\Big)^2\Big]ds+\Delta_l^2\Big).
\end{align*}
On applying Gronwall's lemma, the result is concluded.

\end{proof}

\subsection{Results for the i.i.d.~Multilevel DEnBKF}\label{app:ml_res}

In order to prove  \autoref{theo:main_theo} we require the following two results.

\begin{prop}\label{prop:var_term1}
{For any $T\in\mathbb{N}$ fixed and $t\in[0,T-1]$ there exists a $\mathsf{C}<+\infty$ such that for any $(l,N,k_1)\in\mathbb{N}_0\times\{2,3,\dots\}\times\{0,1,\dots,\Delta_l^{-1}\}$:
$$
\mathbb{E}\Big[\Big\|[\hat{\eta}_{t+k_1\Delta_l}^{N,l}-\eta_{t+k_1\Delta_l}](e)\Big\|_2^2\Big] \leq \mathsf{C}\Big(\frac{1}{N}+\Delta_l^2\Big).
$$}
\end{prop}

\begin{proof}
The proof follows similarly to that in \cite{CJY20}, where we use a combination of both the $C_2$--inequality, Jensen's inequality
and the strong error attained in  \autoref{lem:strong_error}.
\end{proof}

\begin{prop}\label{prop:var_term2}{
For any $T\in\mathbb{N}$ fixed and $t\in[0,T-1]$ there exists a $\mathsf{C}<+\infty$ such that for any $(l,N,k_1)\in\mathbb{N}\times\{2,3,\dots\}\times\{0,1,\dots,\Delta_{l-1}^{-1}\}$:
$$
\mathbb{E}\Big[\Big\|[\hat{\eta}_{t+k_1\Delta_{l-1}}^{N,l}-\hat{\eta}_{t+k_1\Delta_{l-1}}^{N,l-1}](e) - [\eta_{t+k_1\Delta_{l-1}}^{l}-\eta_{t+k_1\Delta_{l-1}}^{l-1}](e)\Big\|_2^2\Big] \leq \frac{\mathsf{C}\Delta_l}{N}.
$$}
\end{prop}

\begin{proof}
 One can use standard properties of i.i.d. empirical averages, along with the strong error attained in  \autoref{lem:strong_error}.
\end{proof}

\end{document}